\documentclass[aps,prc,twocolumn,superscriptaddress,preprintnumbers]{revtex4-2}
\usepackage[utf8]{inputenc}
\usepackage{physics}
\usepackage{slashed}
\usepackage{amssymb}
\usepackage{cancel}
\usepackage{xcolor}
\usepackage[inline]{enumitem}
\usepackage{comment}
\newcommand\numberthis{\addtocounter{equation}{1}\tag{\theequation}}
\def\ba#1\ea{\begin{align*}#1\numberthis{}\end{align*}}
\newcommand{\be}{\begin{equation}}
\newcommand{\ee}{\end{equation}}
\newcommand{\kf}{p_\text{F}}

\newcommand{\reb}[1]{\left[{#1}\right]}
\newcommand{\rob}[1]{\left({#1}\right)}
\newcommand{\tb}[1]{\left<{#1}\right>}
\newcommand{\f}[2]{\frac{#1}{#2}}
\newcommand{\bs}[1]{\boldsymbol{#1}}

\newcommand{\de}{\delta}

\usepackage[sep=4pt, offset=1.em]{simpler-wick}
\usepackage{mathtools}

\newcommand{\q}[1]{{Q}_{{#1}}}
\newcommand{\qbar}[1]{\bar{Q}_{{#1}}}
\newcommand{\p}[1]{{P}_{{#1}}}
\newcommand{\opt}[1]{\hat{T}_{{#1}}}
\newcommand{\opti}[1]{\hat{\tilde{T}}_{{#1}}}
\newcommand{\opty}[1]{\hat{T}^\infty_{{#1}}}

\newcommand{\optd}[1]{\Delta\hat{T}^{{#1}}}
\newcommand{\Vop}{\hat{V}}
\newcommand{\Gop}{\hat{G}_0}

\newcommand{\Sop}{{\cal\hat{S}}}

\newcommand{\avec}{{\bs{a}}}
\newcommand{\bvec}{{\bs{b}}}

\newcommand{\ivec}{{\bs{i}}}
\newcommand{\jvec}{{\bs{j}}}
\newcommand{\kvec}{{\bs{k}}}
\newcommand{\lvec}{{\bs{l}}}

\newcommand{\xvec}{{\bs{x}}}

\newcommand{\rvec}{{\bs{r}}}
\newcommand{\svec}{{\bs{s}}}
\newcommand{\pvec}{{\bs{p}}}
\newcommand{\qvec}{{\bs{q}}}

\newcommand{\Pvec}{{\bs{P}}}

\newcommand{\nph}[1]{$#1p#1h$}
\def\({\left(} 
\def\){\right)}
\def\[{\left[} 
\def\]{\right]}
\begin{document}

\preprint{LA-UR-22-31633}

% =================================
% =================================
%           TITLE
% =================================
% =================================
\title{On nuclear short-range correlations and the zero-energy eigenstates of the Schr{\"o}dinger equation}

\author{Saar Beck}
\affiliation{The Racah Institute of Physics, The Hebrew University, 
Jerusalem 9190401, Israel}
\author{Ronen Weiss}
\affiliation{Theoretical Division, Los Alamos National Laboratory, Los Alamos, New Mexico 87545, USA}
\author{Nir Barnea}
\affiliation{The Racah Institute of Physics, The Hebrew University, 
Jerusalem 9190401, Israel}

\date{\today}
% =================================
% =================================
%           ABSTRACT
% =================================
% =================================
\begin{abstract}
We present a systematic analysis of the nuclear 2 and 3-body short range 
correlations, and their relations to the zero-energy eigenstates of the Schr{\"o}dinger equation. 
To this end we analyze the doublet and triplet Coupled-Cluster amplitudes in the
high momentum limit, and show that 
they obey universal equations independent of the number of nucleons and their state.
Furthermore, we find that
these Coupled-Cluster amplitudes coincide with the zero-energy
Bloch-Horowitz operator.
These results illuminate the relations between the nuclear many-body theory and the generalized contact formalism, introduced to describe the nuclear 2-body short range correlations, and it might also 
be helpful for general Coupled-Cluster computations as the asymptotic part of the amplitudes is given and shown to be universal. 
\end{abstract}
\maketitle
% =================================
% =================================
%           INTRODUCTION
% =================================
% =================================
\section{Introduction}
Nuclear short-range correlations (SRC) have been studied
extensively over the last few decades (see Refs. \cite{Atti:2015eda,Hen:2016kwk} for recent reviews). Large momentum-transfer quasi-elastic
electron and proton scattering reactions are the main experimental tools 
facilitating these studies \cite{Frankfurt81,Frankfurt88}. In such reactions, interpreted in a high
resolution picture, back-to-back SRC nucleon pairs were clearly identified \cite{tang03,piasetzky06,shneor07,subedi08,korover14,Cohen:2018gzh},
with a significant dominance of 
neutron-proton pairs \cite{hen14,Duer:2018sxh,korover14,schmidt20,Korover:2021}.
Inclusive reactions where used to study the abundance of such SRC pairs  \cite{frankfurt93,egiyan02,egiyan06,fomin12,Schmookler:2019nvf}.
Currently, ab-initio approaches are unable to directly calculate the cross sections
of these reactions, in all but the lightest nuclei. Nevertheless, qualitatively similar conclusions were obtained in structure studies,
that focused mainly on the high momentum
tail of the nuclear momentum distribution \cite{Feldmeier:2011qy,Alvioli:2012qa,wiringa14,Rios:2013zqa,Ryckebusch:2019oya,Alvioli:2013qyz,wiringa14,neff15,ryckebusch15,Alvioli:2007zz}.
The study of nuclear three-body SRCs, i.e. three nucleons at close proximity,
is still very preliminary at this stage \cite{Ye18,Sargasian19} and their impact on nuclear quantities is still mostly unknown.

Following Tan's work on ultra-cold atoms \cite{Tan08a,Tan08b,Tan08c,Braaten12},
the Generalized Contact Formalism (GCF) was introduced and utilized
to analyze SRC effects in nuclei \cite{Weiss14,Weiss:2015mba,Weiss_2016,Weiss17_CoupledChannels}. It is based on the scale separation ansatz,
assuming a factorization of the nuclear wave-function when two nucleons are close to each other. The GCF provides a framework to study both nuclear structure and nuclear reactions, and
was successfully tested against ab-initio studies, providing a good description
of both two-body densities at short distance and high-momentum tails of different momentum distributions \cite{Weiss:2015mba,Weiss:2016obx,Cruz-Torres2020}. In addition, the GCF is found to be in good agreement with
exclusive electron scattering experiments and other reactions sensitive to SRC pairs \cite{Weiss_2016,Weiss_EPJA16,Weiss:2018tbu,schmidt20,WEISS2019484,
Duer:2018sxh,schmidt20,Pybus:2020itv,Korover:2021,patsyuk2021unperturbed}.
As such, the GCF allows for a quantitative comparison between
ab-initio calculations and experimental results, with direct connection to the underlying nuclear interaction. The GCF results lead to a comprehensive and consistent picture
of nuclear SRCs, where the only tension is with respect to the analysis of inclusive reactions \cite{weiss2020inclusive}. Recently, shell-model calculations were combined with the GCF to calculate nuclear matrix elements for neutrinoless double beta decay \cite{weiss2021neutrinoless},
taking into account both short-range and long-range contributions consistently.  

As pointed out, the GCF is based on the asymptotic factorization ansatz for the many-body nuclear
wave-function $\Psi$, when nucleon $i$ is close to nucleon $j$ \cite{Weiss:2015mba}
\be
  \Psi\xrightarrow[r_{ij}\rightarrow 0]{}
  \sum_\alpha\varphi_{ij}^\alpha\left(\bs{r}_{ij}\right)
  A_{ij}^\alpha\left(\bs{R}_{ij},\{\bs{r}_k\}_{k\not=i,j}\right).
\ee
In this picture, particles $i$ and $j$ are strongly interacting, and, therefore,
described by a two-body function $\varphi_{ij}^\alpha$, decoupled from the
rest of the system, which is described by the function $A_{ij}^\alpha$.
In the GCF,  $\varphi_{ij}^\alpha$ is assumed to be universal, i.e. independent
of the nucleus or its many-body state, and is defined to be the zero-energy solution
of the two-body Schr\"{o}dinger equation with quantum numbers $\alpha$, obtained with the same
nucleon-nucleon interaction model used for the many-body wave-function.
A similar factorization should hold in momentum space, for pairs with
high relative momentum $\bs{k}_{ij}$ 
\begin{align} \label{asymp_k}
\tilde{\Psi}(\bs{k}_1,\bs{k}_2,...,\bs{k}_A)
\xrightarrow[k_{ij}\rightarrow \infty]{}
\sum_\alpha \tilde{\varphi}_{ij}^\alpha\left(\bs{k}_{ij}\right)
\tilde{A}_{ij}^\alpha\left(\bs{K}_{ij},\{\bs{k}_n\}_{n\not=i,j}\right),
\end{align}
where $\tilde{\varphi}_{ij}^\alpha$ and $\tilde{A}_{ij}^\alpha$
are respectively the Fourier transforms of $\varphi_{ij}^\alpha$ and $A_{ij}^\alpha$.
Based on these asymptotic factorizations, nuclear contact matrices are defined as 
\be
C_{ij}^{\alpha \beta} = N_{ij}\langle A_{ij}^\alpha | A_{ij}^\beta \rangle.
\ee
Here, $ij$ stands for one of the pairs: proton-proton, neutron-neutron or neutron-proton,
and $N_{ij}$ is the total number of $ij$ pairs in the nucleus. The diagonal contact elements
$C_{ij}^{\alpha \alpha}$ are proportional to the number of SRC pairs with qunatum number
$\alpha$ in a given nuclear state. 

The asymptotic factorization, including the definition of the universal
two-body functions, is the underlying assumption for
the GCF predictions, and was verified numerically using ab-initio calculations \cite{Weiss:2015mba,Weiss:2016obx,Cruz-Torres2020}.
It is also supported by the work of Refs. \cite{Anderson2010,Bogner12,Tropiano2021}, based on renormalization group
arguments. 
In view of its success, the two-body GCF is expected to be the leading 
order term of a short-range (or a high-momentum) 
expansion of the nuclear wave-function.
However, next order corrections are currently not well understood, especially 
the role of the elusive SRC triplets. 

In this work we study the asymptotic form of the nuclear wave-function
using the Coupled Cluster (CC) expansion \cite{Bartlett07,Hagen_2014}, aiming  to put the GCF on a more solid theoretical grounds. 
In addition, the CC expansion provides a systematic way to include higher order corrections, e.g. 3-body SRCs, beyond the leading 2-body SRC term of the asymptotic expansion of the many-body wave-function.
Here, we limit our attention to Hamiltonians containing only 2-body interaction, postponing the discussion of 3-body forces to future works.  

The paper is organized as follows. In section \ref{sec:cc} we provide 
a short introduction to the CC expansion method. Then, in section \ref{sec:bss} we discuss the momentum basis and its merits. 
The derivation of the high-momentum asymptotic equations governing the behavior
of two-body and three-body SRCs is presented in section \ref{sec:cc amp high}.
In section \ref{sec:2b} we focus on
two-body correlations and analyze their universal behavior. 
Three-body effects are then analyzed in section \ref{sec:3b}, where we derive
the appropriate universal equation and show its relation to the solution
of the zero-energy three-body problem.
For the sake of brevity some more technical details are presented in the appendix.

%=============================================================================
%       THE COUPLED CLUSTER THEORY
%=============================================================================
\section{Coupled cluster theory}\label{sec:cc}
The general form of a Hamiltonian describing a many-particle system interacting via  two-body potential $\hat{V}$ is given by
\begin{align}
   &\hat{H} \equiv \hat{H}_0+\hat{U}+\Vop 
   \cr &= 
    \sum_{r} \epsilon_r \bs{r}^\dagger\bs{r} 
    +\sum_{rr'} U^{r}_{r'} \rvec^\dagger\rvec' 
    +\f{1}{4}\sum_{rsr's'}V^{rs}_{r's'}\bs{r}^\dagger\bs{s}^\dagger\svec'\rvec',
\end{align}
where $\hat{H}_0$ is the ``zero-order" or unperturbed Hamiltonian (not necessarily the free Hamiltonian), and 
$\hat{U}$ is the residual one-body interaction.
The operators $\rvec,\svec,\ldots$ are the usual fermionic ladder operators 
corresponding to the single particle eigenstates $\ket{r},\ket{s},\ldots$ of $\hat{H}_{0}$, 
i.e. $\hat{H}_{0} \ket{{r}} = \epsilon_r \ket{{r}}$,
or equivalently
\be \label{eq:H0com}
    [\hat{H}_{0},\rvec^\dagger]=\epsilon_r\rvec^\dagger
    \qquad
    [\hat{H}_{0},\rvec]=-\epsilon_r\rvec,
\ee
where $[\hat{A},\hat{B}]$ is the regular commutator.
They obey the anti-commutation relations
\be \label{eq:anticom}
   \{\rvec,\svec\}=0,
   \hspace{2em}
   \{\rvec^\dagger,\svec^\dagger\}=0,
   \hspace{2em}
   \{\rvec^\dagger,\svec\} = \de_{rs}.
\ee

In the following we will use the notation $\ket{r_1r_2\ldots r_A}$ to denote
normalized antisymmetrized $A$-body states and $|r_1r_2\ldots r_A)$ to denote the simple, non-symmetrized, many-body states, e.g. 
$\ket{r s} = \frac{1}{\sqrt{2}}[|r s)-|s r)]$.
The matrix elements of the 2-body potential $\Vop$
are then given by
\be
   V^{rs}_{r's'} = \bra{r s}\hat{V}\ket{ r's'}
   = (r s | \Vop | r's')-(rs|\Vop|s'r').
\ee

The starting point of the CC method is a reference Slater-determinant state $\ket{\Phi_0}$,
composed of $A$ single particle states.
In general, a wave function $|\Psi\rangle$ is a linear combination of all such Slater determinants. These determinants can be organized in a systematic way, by considering first the determinants obtained replacing a state occupied in $\ket{\Phi_0}$ with a state not occupied in $\ket{\Phi_0}$, than replacing two such states, and so on.
Following the convention of Shavitt \& Bartlett \cite{shavit2009}, we use the letters $i,j,\ldots,n$ to denote "hole" states, i.e. single-particle states that are occupied in $|\Phi_0\rangle$, and the letters $a,b,\ldots,f$ to denote "particle" states, i.e. single-particle states that are not occupied in $|\Phi_0\rangle$. $r,s,\ldots,w$ will be used to denote both states. Therefore,
\ba
    \bs{i}^\dagger \ket{\Phi_0} = 0\;,
    \quad \text{and} \quad
    \bs{a} \ket{\Phi_0} =0.
\ea
The interacting many-body state $|\Psi \rangle$, an eigenstate of $\hat{H}$, is written in the CC formulation as
\be\label{Psi_CC}
  \ket{\Psi} = e^{\opt{}} \ket{\Phi_0},
  \quad\text{where}\quad
  \opt{} = \sum_n \opt{n},
\ee
and
\be
  \opt{n} = \f{1}{n!^2}\sum_{a_1 \ldots a_n, i_1 \ldots i_n}
    t^{a_1 a_2\ldots a_n}_{i_1 i_2 \ldots i_n} \bs{a}_1^\dagger\bs{a}_2^\dagger\cdots
                                      \bs{i}_2\bs{i}_1 
\ee
is the $n$-particle, $n$-hole (\nph{n}) cluster operator.

To determine the amplitudes $t^{a_1 a_2\ldots a_n}_{i_1 i_2 \ldots i_n}$, a set of non-linear equations, the CC equations, can be obtained by projecting the Schr{\"o}dinger equation on an \nph{n} state $\ket{\Phi^{ab\cdots}_{ij\cdots}}\equiv
  \avec^\dagger\bvec^\dagger\cdots\jvec\ivec \ket{\Phi_0}$.
The full derivation of the CC equations is given, e.g., in Ref. \cite{shavit2009}.
Omitting the 1-body potential term $\hat{U}$ and the \nph{1} cluster operator $\opt{1}$, the two- and three-body CC equations are given by
\begin{align}
  \label{eq:fullteq2}
  0=\bra{\Phi_{ij}^{ab}}
  \Vop&+[\hat{H}_{0},\opt{2}]+ [\Vop,\opt{2}]
  +\f{1}{2}[[\Vop,\opt{2}],\opt{2}]
  \cr &
   + [\Vop,\opt{3}]+[\Vop,\opt{4}]
  \ket{\Phi_0},
  \\
  \label{eq:fullteq3}
  0=\bra{\Phi_{ijk}^{abc}}&
  [\hat{H}_{0},\opt{3}]
   + [\Vop,\opt{2}]
  + \f{1}{2}[[\Vop,\opt{2}],\opt{2}]+[\Vop,\opt{3}]
  \cr&+ [[\Vop,\opt{2}],\opt{3}]+[\Vop,\opt{4}]
  + [\Vop,\opt{5}]
  \ket{\Phi_0}.
\end{align}

%===========================================================
% Basis states and their implications
%===========================================================
\section{Momentum basis states}\label{sec:bss}

To study SRCs it is most convenient to work with single-particle basis states, i.e. the eigenstate of $\hat{H}_0$, that have well defined momentum. 
This choice is natural for an infinite system, like nuclear matter - see e.g. \cite{CCNM2013,CCNM2014}, but it might seem rather odd for describing a bound nucleus which is a compact object. However, large nuclei have relatively constant density and far from the surface behave like an infinite nuclear system.
Thus, we set the problem in a box of size $L$ with periodic boundary conditions. For $L$ larger than the nucleus size, the wave-function and the binding energy
approach very fast the free space ($L\to \infty$) values and we need not
worry about the impact of the boundary conditions on the nuclear SRCs.

Assuming $\pvec = (p_1,p_2,p_3)$ to be a triad of integers, the basis states $\{\ket{\pvec}\}$ 
\be
   \bra{\xvec}\ket{\pvec}=\frac{1}{\sqrt{\Omega}}e^{i\frac{2\pi}{L}\xvec\cdot \pvec},
   \quad
   \bra{\pvec}\ket{\pvec'}=\delta_{\pvec,\pvec'},
\ee
with $\Omega=L^3$, is a complete set of orthonormal states, which combined with the spin and isospin degrees of freedom form our single-particle basis.
A natural choice for $\ket{\Phi_0}$, the starting point of the CC expansion, is a Slater determinant composed of the $A$ lowest kinetic energy single-particle states.

If there is a well defined Fermi momentum $p_F$,
such that all the hole states are momentum states with momentum smaller than
$p_F$, while particle states have momentum larger than $p_F$,
then the system is called a \textit{closed shell} system. 
To simplify matters, in the following we shall restrict our the discussion to such closed shell systems only.

Working with this single-particle momentum basis, $\hat{H}_0$ coincides with the kinetic energy operator and therefore $\hat{U}=0$.
The Slater-determinant  $\ket{\Phi_0}$, as well as the
$npnh$ states $\ket{\Phi^{ab\cdots}_{ij\cdots}}$, are a product of single-particle momentum states, hence they are eigenstates of the total center of mass (CM) momentum operator $\hat{\Pvec}_{CM}$.
The two-body potential is translational invariant, hence the CM momentum is a good quantum number, and the wave-function $\ket{\Psi}$ is also an eigenstate of $\hat{\Pvec}_{CM}$,
\ba\label{Pcm}
   \hat{\Pvec}_{CM}\ket{\Psi}
 =\hat{\Pvec}_{CM}e^{\opt{}}\ket{\Phi_0}
   = \Pvec_{CM}\ket{\Psi}.
\ea
Closing the last equation with $\bra{\Phi_0}$ and acting with $\hat{\Pvec}_{CM}$ once to the left and once to the right, and noting 
that $\bra{\Phi_0}\ket{\Psi}\neq 0$, we must conclude that
$\ket{\Phi_0}$ and $\ket{\Psi}$ share the same eigenvalue of the total momentum  $\Pvec_{CM}$.

We may now repeat the same argument for the \nph{1} states. 
Closing Eq. \eqref{Pcm} with $\bra{\Phi_i^a}$ and using 
$\langle \Phi_i^a | \Psi \rangle = t_i^a$ we get
\be
   (\pvec_a-\pvec_i)t^a_i=\bs{0},
\ee
which for all closed shell systems implies \cite{Bishop78,CCNM2013}
\be
   t^a_i=0,
\ee
because $(\pvec_a-\pvec_i)\neq\bs{0}$, as $\pvec_a$ corresponds to a particle state while $\pvec_i$ to a hole state.
Thus, with this choice of basis states, $\opt{1}$ is eliminated from the CC
expansion, as was assumed in Eqs. \eqref{eq:fullteq2} and \eqref{eq:fullteq3}.

Considering now the \nph{2} states, multiplying Eq.
\eqref{Pcm} by $\bra{\Phi^{ab}_{ij}}$ one gets \cite{CCNM2014}
\be
  (\pvec_a+\pvec_b-\pvec_i-\pvec_j)t^{ab}_{ij}=\bs{0}.
\ee
This implies that $\opt{2}$ conserves momentum, i.e. $t_{ij}^{ab}=0$
if $\pvec_a+\pvec_b-\pvec_i-\pvec_j \neq \bs{0}$.
It can be similarly shown that for a closed shell system all amplitude operators 
$\opt{n}$ must conserve momentum.

%===========================================================
% Coupled cluster amplitudes in the high momentum limit 
%===========================================================
\section{Coupled cluster amplitudes in the high momentum limit }\label{sec:cc amp high}
SRCs are associated with high momentum particles. To understand their role in the many-body wave-function we need to study the high momentum behavior of the CC amplitudes $\opt{n}$ as dictated by Eqs. \eqref{eq:fullteq2} and \eqref{eq:fullteq3}. In the following we will assume $a,b$ and $c$ to be highly excited states corresponding to momenta $p_a,p_b,p_c\gg\kf$. We note that in order for the wave-function to be properly normalized the CC amplitudes $\opt{n}$ must vanish in this limit, 
e.g. $t^{abc}_{ijk}\to 0$ when 
$a,b,c\to\infty$.

For a system of fermions, we expect the CC amplitudes to admit the natural hierarchy, where double excitations are much more significant than three-body excitations which on their part are more important than the four-body excitations, etc.
It follows that the contributions of $[\hat{V},\opt{3}]$ and $[\hat{V},\opt{4}]$ to the 2-body equation can be neglected.
Similarily, the terms $[\hat{V},\opt{4}]$ and $[\hat{V},\opt{5}]$ can be neglected in the 3-body CC equation.

In order to understand the behaviour of the CC amplitudes in the high momentum limit, let us inspect the $\opt{2}$ equation, Eq. \eqref{eq:fullteq2}, in the limit $p_a,p_b\to\infty$.
In this case, the leading terms are the source term $V^{ab}_{ij}$ and the kinetic energy term $[\hat{H}_0,\opt{2}]$. Retaining only these terms 
leads to the well known asymptotic result
\be\label{t2faraway}
 t^{ab}_{ij} \rightarrow -\frac{1}{E^{ab}_{ij}}V^{ab}_{ij},
\ee
were $E^{ab}_{ij}$ is the excitation energy given by the relation
\be
 E^{a_1a_2\ldots a_n}_{i_1 i_2\ldots i_n}\equiv 
 (\epsilon_{a_1}+\epsilon_{a_2}+\ldots \epsilon_{a_n})
 -(\epsilon_{i_1}+\ldots+\epsilon_{i_n}).
\ee

If, for $p_a,p_b\to\infty$, the potential matrix elements $V^{ab}_{ij}$ are independent of the exact holes states, 
i.e. $V^{ab}_{ij}\approx V^{ab}_{0_i0_j}$, with $0_i$ being a zero-momentum state (used loosely to indicate the lowest momentum state with the same quantum numbers as the state $i$), the asymptotic two-body amplitude presented in Eq. \eqref{t2faraway} is universal in
the limited sense. That is, $t^{ab}_{ij}\approx-\frac{1}{E^{ab}_{00}}V^{ab}_{00}$ is independent of the number of nucleons $A$ and the specifics of the nuclear state. On the other hand, it depends on the potential - therefore its universality is limited.
This form of asymptotic behaviour was first suggested by Amado \cite{amado76}
exploring the asymptotic form of the nuclear momentum distribution. 
It turns out, however, that although Eq. \eqref{t2faraway} is asymptotically correct, it is valid only at extremely high momentum, larger than 10 fm$^{-1}$, making it impractical for actual calculations \cite{Zabolitzky78}. Consequently, in order to get a reasonable description of the asymptotic nuclear wave-function, we must retain more terms besides
the source term and the $[\hat{H}_0,\opt{n}]$ terms in the CC equations.

With the 3, and 4-body amplitudes neglected, the CC $\opt{2}$ equation, Eq. \eqref{eq:fullteq2}, takes the form
\be  \label{eq:teq2p}
  0=\bra{\Phi_{ij}^{ab}}
  \Vop+[\hat{H}_{0},\opt{2}]+ [\Vop,\opt{2}]
  +\f{1}{2}[[\Vop,\opt{2}],\opt{2}]
  \ket{\Phi_0}.
\ee
Comparing now the terms $[\Vop,\opt{2}]$, and $[[\Vop,\opt{2}],\opt{2}]$ we note that for the latter we get the following matrix elements, ignoring  combinatorical factors, 
\be\label{eq:t2t2}
  V^{kl}_{de}t^{ab}_{ik}t^{de}_{jl}, \quad
  V^{kl}_{de}t^{ab}_{kl}t^{de}_{ij}, \quad
  V^{kl}_{de}t^{ad}_{ik}t^{be}_{jl}, \quad
  V^{kl}_{de}t^{ad}_{ij}t^{be}_{kl}.
\ee
Here, for brevity, we use the Einstein convention assuming implicit summation on repeated lower and upper indices.
In the high momentum $p_a,p_b\to \infty$ and low density
$i,j,k,l\to 0$ limit, these terms take the form
\be\label{eq:t2kf0}
  2V^{00}_{de}t^{de}_{00}t^{ab}_{00}, \quad \text{and} \quad
  2V^{00}_{de}t^{ad}_{00}t^{be}_{00}.
\ee
The first of these terms is nothing but an energy shift, a correction to the 
excitation energy $E^{ab}_{ij}$, which we can neglect in the high momentum limit. 
%Notating $-a$ for a single particle state identical to $a$ but with opposite momentum, i.e. $\pvec_{-a}=-\pvec_a$,
We note that the second term is zero unless
%$d=-a$, $e=-b$, and $b=-a$.
$\pvec_d=-\pvec_a$, $\pvec_e=-\pvec_b$ and $\pvec_b=-\pvec_a$.
This term is clearly suppressed by a factor of $t^{ab}_{00}$ with respect to   $[\Vop,\opt{2}]$, 
and thus can be neglected as well.
 
Considering now the $\opt{3}$ equation, Eq. \eqref{eq:fullteq3}. After neglecting the $\opt{4}$, $\opt{5}$ as well as the $[[\Vop,\opt{2}],\opt{3}]\ll[[\Vop,\opt{2}],\opt{2}]$ terms, we remain with 
\ba
  0 = \bra{\Phi_{ijk}^{abc}}&
  [\hat{H}_{0},\opt{3}]+ [\Vop,\opt{2}]
  \cr&
  + \f{1}{2}[[\Vop,\opt{2}],\opt{2}]+[\Vop,\opt{3}]
  \ket{\Phi_0}. 
\ea
Comparing again the $[\Vop,\opt{2}]$, and $[[\Vop,\opt{2}],\opt{2}]$, we see that the only terms that survive in the high momentum/low density limit are respectively $V^{ab}_{e0}t^{ce}_{00}$, and $V^{a0}_{ef}t^{be}_{00}t^{cf}_{00}$. 
Thus as before, the double commutator term is suppressed by a factor of $t^{ab}_{00}$ and can be neglected.

Summing up, in the limit of high momenta, we expect the two- and three-body CC amplitudes to obey the equations: 
\begin{align}
  \label{eq:t2asym}
  0 &= \bra{\Phi_{ij}^{ab}} [\hat{H}_{0},\opt{2}]+[\Vop,\opt{2}]+\Vop
  \ket{\Phi_0},
  \\
  \label{eq:t3asym}
  0 &= 
  \bra{\Phi_{ijk}^{abc}} [\hat{H}_{0},\opt{3}]
   + [\Vop,\opt{3}]+[\Vop,\opt{2}] \ket{\Phi_0}. 
\end{align}
In the following sections we will analyze these equations.

%=============================================================================
% THE 2-BODY AMPLITUDES IN THE HIGH MOMENTUM LIMIT
%=============================================================================
\section{The 2-body amplitude}\label{sec:2b}
In section \ref{sec:cc amp high} we argued that asymptotically the 2-body CC equation takes the form of Eq. \eqref{eq:t2asym}. 
In order to evaluate this equation, we note that there can be no contractions between the operators $\bs{a}^\dagger\bs{b}^\dagger \bs{j}\bs{i}$ that appear in the bra state, hence all the labels $a,b,i$ and $j$ has to appear on the amplitude and potential operators. Therefore
$\bra{\Phi_{ij}^{ab}}\Vop\ket{\Phi_0}=V^{ab}_{ij}$, and $\bra{\Phi_{ij}^{ab}}\reb{\hat{H}_{0},\opt{2}}\ket{\Phi_0}=E^{ab}_{ij}t^{ab}_{ij}$. To evaluate the commutator $\reb{\Vop,\opt{2}}=\Vop\opt{2}-\opt{2}\Vop$, we note that all the operators of $\Vop$ in $\Vop\opt{2}$ can be moved to the right of $\opt{2}$ and then the term will cancel with $\opt{2}\Vop$.
In the process, all possible contractions between $\Vop$ and $\opt{2}$ will arise, i.e. at least one contraction should be made between them. This commutators yield 5 distinct terms, that combined with the potential and the $\hat{H}_0$ terms
% and $\hat{U}$ (for which we can use $V^{wx}_{yz}\to U^{w}_y\delta^{x}_z$ when there is no internal summation)
results in the linear Coupled-Cluster doublets (CCD) equation% (calculation of the combinatorical factors is given in appendix \ref{sec:num_fac})
\ba\label{eq:t2me}
  0 &= V^{ab}_{ij} + E^{ab}_{ij}t^{ab}_{ij} 
  + \f{1}{2}V^{ab}_{de}t^{de}_{ij}
  \cr &
  + \f{1}{2}V^{kl}_{ij}t^{ab}_{kl}
  + V^{ak}_{id}t^{bd}_{jk}
  + V^{ak}_{kd}t^{bd}_{ij}
  + V^{lk}_{ik}t^{ab}_{jl}
  \cr &
%   +U^k_jt^{ab}_{ik}+U^b_dt^{ad}_{ij}
  + \text{permutations}\;.
\ea
The title `permutations' stands for anti-symmetrization with respect to the indices $ab$ or $ij$ when not placed on the same matrix elements.
 The summation $V^{k v}_{k w}$ is performed only on hole states as the string $\bs{r}^\dagger\bs{s}$ of $\Vop$, where both $\rvec,\svec$ are particle operators, is already normal ordered and therefore its contraction is zero.

In the limit of high momentum/low density the 2nd line of \eqref{eq:t2me}
take either the form $V^{00}_{00}t^{ab}_{00}$ or $V^{a0}_{a0}t^{ab}_{00}$. In both cases these terms enter as small corrections to the excitation energy. 
It follows that these terms can be neglected for large $p_a,p_b$ with respect to the terms appearing on the first line.

Refining this argument, due to momentum conservation we expect that in the limit $p_a,p_b\to\infty$ all the $\opt{2}$ terms on the 2nd and 3rd lines of Eq. \eqref{eq:t2me}, will be either exactly or at least approximately equal to $t^{ab}_{ij}$ since the hole states carry only low momentum of the order of $\kf$ and we assume weak
%dependence on the hole states and weak dependence of the quantum numbers on the momenta behaviour of the CC amplitudes.
momenta dependance on the hole states quantum numbers.
For example, in
$V^{ak}_{id}t^{bd}_{jk}$ the momentum $\pvec_d$ associated with the state $d$
must be of the order $\pvec_d=\pvec_a+\order{\kf}$ implying that
$t^{bd}_{jk}\approx t^{ba}_{ij}$.
Comparing these terms to the term $E^{ab}_{ij}t^{ab}_{ij}$, we see that
for large enough excitations 
\be\nonumber 
   E^{ab}_{ij} \gg  \sum_{kl}V^{kl}_{ij},\; \sum_{kd}V^{ak}_{id},\; \sum_{kd}V^{ak}_{kd},\; \sum_{kl}V^{lk}_{ik}
\ee 
and these terms can be neglected. 
It is important to observe that the neglected terms are all intensive and do not scale with the size of the system.
\begin{comment}
In order to evaluate the interaction sums such as $S=\sum_{kl}V^{kl}_{ij}$
it is convenient to study the infinite volume limit with a local potential (depends on the momentum transfer), and use the 
continuous momentum basis
states
\be
   \bra{\xvec}\ket{\pvec}=\frac{1}{\sqrt{(2\pi)^3}}e^{i\xvec\cdot \pvec},
   \quad
   \bra{\pvec}\ket{\pvec'}=\delta^3(\pvec-\pvec').
\ee
In this limit the sum over hole states is replaced by an integral
\be
   S = \sum_{kl}V^{kl}_{ij}\to S=\int^{\kf}\!d\kvec d\lvec 
   \bra{\kvec\lvec} \hat{V} \ket{\bs{i}\bs{j}}.
\ee
For readability, lets ignore the exchange term. The matrix element takes the form
\ba
   \int^{\kf}\hspace{-1em}&d\kvec d\lvec 
   \bra{\kvec\lvec} \hat{V} \ket{\bs{i}\bs{j}}
   \\&=
   \int^{\kf}\hspace{-1em}d\kvec d\lvec
   \delta^3\rob{\bs{k}+\bs{l}-\bs{i}-\bs{j}}
   V\rob{\f{1}{2}\reb{\bs{k}-\bs{l}-\bs{i}+\bs{j}}}
   \\&=
   \int^{\kf}\hspace{-1em}d\kvec
   V\rob{\bs{k}-\bs{i}}
   \\& \leq
   \int^{2\kf}\hspace{-1em}d\qvec
   \abs{V\rob{\bs{q}}}
   =\tb{\abs{V}}\f{4\pi}{3}\rob{2\kf}^3
\ea
which is a finite intensive number, bounded up by the product of the
average of the absolute value of the potential in a $2k_F$ sphere
times the density
which is relatively constant through the nuclear chart. 
\end{comment}

The resulting 2-body amplitude equation is then
\ba\label{eq:t2high}
  0 &= t^{ab}_{ij} 
     + \f{1}{E^{ab}_{ij}}V^{ab}_{ij}
     + \f{1}{2E^{ab}_{ij}}V^{ab}_{de}t^{de}_{ij}
   \;,
\ea
which is nothing but particle-particle ladder approximation of the CCD equation, applied for example in Ref. \cite{CCNM2013} to estimate the nuclear matter equation of state.
In the following we will use the notation $\opty{2}$ to denote the solution of Eq. \eqref{eq:t2high} in the non-symmetrized basis with $E^{ij}\equiv \epsilon_i + \epsilon_j \to 0$. As it is a linear equation, $\opty{2}$ is unique. We will show in appendix \ref{sec:t2asym} that indeed 
$\bra{\bs{ab}}\opt{2}\ket{\bs{ij}}\to\bra{\bs{ab}}\opty{2}\ket{\bs{ij}}$ as $p_a,p_b\to\infty$.

We can now discuss the properties of $\opty{2}$.
%and its implications to the factorization of the nuclear wave-function. 
As stated above, the cluster operator $\opty{2}$ is defined as the solution of the equation
\ba\label{eq:t2univ}
  0 &= \rob{t^\infty}^{ab}_{ij} 
     + \f{1}{E^{ab}}V^{ab}_{ij}
     + \f{1}{2E^{ab}}V^{ab}_{de}\rob{t^\infty}^{de}_{ij},
\ea
where $E^{ab}=\epsilon_a+\epsilon_b$.
A similar result was derived by Zabolitzky in Ref. \cite{Zabolitzky78}.
To analyze this equation, it is convenient to introduce particle-particle and the hole-hole projection operators, 
\be
  \q{2}=\sum_{de}|{d}{e})({d}{e}|,
  \quad 
  \p{2}=\sum_{lm}|{l}{m})({l}{m}|, 
\ee
and the Green's function
\be
\Gop(E)=\frac{1}{E-\hat{H}_{0}+i\varepsilon}.
\ee
Eq. \eqref{eq:t2univ} can then be written as
\be
    \opty{2}=\q{2}\Gop(0)\Vop\opty{2}+\q{2}\Gop(0)\Vop\p{2},
\ee
and formally solved to yield
\be\label{eq:t2infsol0}
   \opty{2}=\f{1}{1-\q{2}\Gop(0)\Vop}\q{2}\Gop(0)\Vop\p{2}\;.
\ee
Clearly, 
$\p{2}\opty{2}=\opty{2}\q{2}=0$ as expected from a cluster operator.
Using the relation $\q{2}\hat{H}_0\p{2}=0$ the solution \eqref{eq:t2infsol0} can be rewritten as (see appendix \ref{sec:t2_to_BH}) 
\ba\label{eq:t2infsol}
   \opty{2} &= 
%   \frac{1}{1-\Gop(0)\q{2}\Vop} \Gop(0)\q{2}\hat{H}\p{2}
%   \cr &=
     \frac{1}{\q{2}(0+i\varepsilon-\hat{H})\q{2}}\q{2}\hat{H}\p{2}
\;.
\ea
Before proceeding, we note that $\p{2}$ is not eqivalent to $\qbar{2}=1-\q{2}$ the complement of $\q{2}$, as $\qbar{2}$ must include not only hole-hole states but also particle-hole states. 
For infinite nuclear matter we expect however, that $\opty{2}$ is translational invariant and therefore we can consider only pairs with zero CM momentum. For such pairs, there are no particle-hole contributions and we can replace $\p{2}$ by $\qbar{2}$.
In this subspace
\be\label{eq:t2infsol2}
   \opty{2} = \frac{1}{\q{2}(0+i\varepsilon-\hat{H})\q{2}}\q{2}\hat{H}\qbar{2}
   \;.
\ee

Comparing now Eq. \eqref{eq:t2infsol2} with the Bloch-Horowitz equations \cite{BH58},
\begin{align}\label{BHqbar2}
   \qbar{2} |\Psi\rangle &= \frac{1}{\qbar{2}(E+i\varepsilon-\hat{H})\qbar{2}}
   \qbar{2} \hat{H} \q{2} |\Psi\rangle
   \\ \label{BHq2}
   \q{2} |\Psi\rangle &= \frac{1}{\q{2}(E+i\varepsilon-\hat{H})\q{2}}
   \q{2} \hat{H} \qbar{2} |\Psi\rangle\;,
\end{align}
it is clear that $\opty{2}$ is nothing but the zero-energy 2-body Bloch-Horowitz operator
\be
    \hat{O}_2^\text{B.H.} = \frac{1}{\q{2}(0+i\varepsilon-\hat{H})\q{2}}
   \q{2} \hat{H} \qbar{2} .
\ee
This operator fulfills the relation $\hat{O}_2^\text{B.H.}\ket{\Psi_2}=\q{2}\ket{\Psi_2}$ 
for any zero energy eigenstate $\ket{\Psi_2}$ that obeys $\hat{H}\ket{\Psi_2}=0$.
It follows that if $\hat{H}\ket{\Psi_2}=0$ and $\hat{\bs{P}}_{CM}\ket{\Psi_2}=0$, then
\be\label{eq:t2inf equiv eigen}
    \q{2}\ket{\Psi_2} = \opty{2}\ket{\Psi_2}. 
\ee 
\begin{comment}
Projecting on the left with $(\bs{ab}|$ we get 
\ba
(\bs{ab}|\psi_{ij})
&=
(\bs{ab}|\opty{2}|\psi_{ij})
=
\sum_{\bs{kl}}(\bs{ab}|\opty{2}|\bs{kl})(\bs{kl}|\psi_{ij})
\\&\approx 
(\bs{ab}|\opty{2}|\bs{ij})\sum_{\bs{kl}}(\bs{kl}|\psi_{ij})
\ea
as claimed. 
\end{comment}
Inspecting eqs. \eqref{eq:t2infsol} and \eqref{eq:t2inf equiv eigen}, we can conclude that 
\begin{enumerate*}[label=(\itshape\roman*)]
 \item The asymptotic 2-body behavior of $\opt{2}$, and therefore of the many-body wave-function, is related to the zero-energy solutions of the 2-body problem.
 \item The relation to the zero-energy solutions show the universality of the asymptotic behavior in the limited sense, that it does not depend on the system, but depends on the potential.
\end{enumerate*}

%=============================================================================
% THE 3-BODY AMPLITUDES IN THE HIGH MOMENTUM LIMIT
%=============================================================================
\section{The 3-body amplitude}\label{sec:3b}
As we have argued in Sec. \ref{sec:cc amp high}, the behavior of the 3-body amplitude $\opt{3}$ at high momentum is dictated by  
Eq. \eqref{eq:t3asym}.
Explicitly, this equation takes the form
\ba\label{eq:t3me}
   0 &= E^{abc}_{ijk}t^{abc}_{ijk} 
   - V^{la}_{ij}t^{bc}_{kl} - V^{ab}_{id}t^{cd}_{jk}
   \cr &
   + \f{1}{2}V^{ab}_{de}t^{cde}_{ijk}
   + \f{1}{2}V^{lm}_{ij}t^{abc}_{klm}
   + V^{al}_{dl}t^{bcd}_{ijk}
   + V^{al}_{id}t^{bcd}_{jkl}
   + V^{lm}_{il}t^{abc}_{jkm}
   \cr &
     + \text{permutations}\;.
\ea
Here, the first term on the rhs is due to $[\hat{H}_0,\opt{3}]$, the next two terms come from the $[\Vop,\opt{2}]$ commutator, and the next five are due to the $[\Vop,\opt{3}]$ commutator. 
The title `permutations' stands for anti-symmetrization with respect to the indices $abc$ or $ijk$ when not placed on the same matrix elements.
Due to momentum conservation, for very large $p_a$ the potential matrix elements $V^{la}_{ij}$ must vanish, leaving $V^{ab}_{id}t^{cd}_{jk}$ as the only source term. 
In addition, all terms coming from the $[\Vop,\opt{3}]$ commutator, except for the first term in the second line (and its corresponding permutations), 
are approximately proportional to $t^{abc}_{ijk}$. 
Therefore, for excitation energy $E^{abc}_{ijk}$ large enough
\be
   E^{abc}_{ijk} \gg
  \sum_{lm}V^{lm}_{ij},\;\sum_{ld}V^{al}_{dl}\;,
  \sum_{ld}V^{al}_{id}\;,
  \sum_{ml}V^{lm}_{il}\;,
\ee
and
the corresponding terms can be neglected in comparison to the free term $E^{abc}_{ijk}t^{abc}_{ijk}$.
As a consequence only the terms $\f{1}{2}V^{ab}_{de}t^{cde}_{ijk}$
remain.
Utilizing these observations, and
defining the symmetrization operator $\Sop_{123}\equiv 1+(123)+(132)$ where $(123)$ is the permutation operator, equation \eqref{eq:t3me} takes the form
\ba\label{eq:t3meeq}
  0 = t^{abc}_{ijk} 
  &+ \frac{\Sop_{abc}(\Sop_{ijk}(V^{ab}_{id}t^{dc}_{jk}))}{E^{abc}_{ijk}}
  + \frac{\Sop_{abc}(V^{ab}_{de}t^{cde}_{ijk})}{2E^{abc}_{ijk}}
  \;.
\ea
% ------------------------------------------------------------------------
%\subsection{Universality \purple{different title?}}

As in the 2-body case we define $\opty{3}$ to be the solution of eq. \eqref{eq:t3meeq} in the limit $E^{abc}_{ijk}\to E^{abc}$, and $t^{dc}_{jk}\to\rob{t^\infty}^{dc}_{jk}$. 
We show in appendix \ref{sec:asymt3} that $\bra{\bs{abc}}\opt{3}\ket{\bs{ijk}}\to\bra{\bs{abc}}\opty{3}\ket{\bs{ijk}}$ as $p_a,p_b,p_c\to\infty$.

To analyze $\opty{3}$ we write  
eq. \eqref{eq:t3meeq} in first quantization using the non-symmetrized basis
defined above.
In the 3-body case, the relation between the anti-symmetrized matrix elements and the non symmetrized ones is
\ba
   \bra{rst}&\hat{O}\ket{uvw}
   \cr &=
   \Sop_{uvw}\[ (rst|\hat{O}|uvw) - (rst|\hat{O}|vuw)\],
\ea
and for a 2-body operator closed by 3-particle states 
\ba
    (rst|\hat{O}_2 & |uvw) \equiv 
    \sum_{i=1}^3 (rst|\hat{O}_2(i)|uvw)
\ea
where $\hat{O}_2(i)$ does not act on the $i$'th particles, e.g. $(rst|\hat{O}_2(3)|uvw)=(rs|{\hat{O}_2}|uv)\delta_{t,w}$.
With the projection operators
\be
  \q{3}=\sum_{def}|def)(def|,
  \quad 
  \p{3}=\sum_{lmn}|lmn)(lmn|, 
\ee
the asymptotic equation for $\opty{3}$ can be written as
\ba\label{eq:t3inf}
   \opty{3}&=\q{3}\Gop\rob{0}\Vop\opty{3}+\q{3}\Gop\rob{0}\Vop\opty{2}\p{3}
   \cr &
   =\f{1}{1-\q{3}\Gop\rob{0}\Vop}\q{3}\Gop\rob{0}\hat{V}\opty{2}\p{3}
   \cr &
   =\f{1}{\q{3}(0+i\varepsilon-\hat{H})\q{3}}\q{3}\hat{H}\opty{2}\p{3}\;.
\ea

Comparing eq. \eqref{eq:t3inf} with the 3-body Bloch-Horowitz equations \cite{BH58} and noting that for 2-body interactions $\q{3}H\qbar{3}=\q{3}H\rob{\q{1}\p{2}+\q{2}\p{1}}$ with $\qbar{3}=1-\q{3}$ 
\begin{align}\label{BHqbar}
   \qbar{3} |\Psi\rangle &= \frac{1}{\qbar{3}(E+i\varepsilon-\hat{H})\qbar{3}}
   \qbar{3} \hat{H} \q{3} |\Psi\rangle
   \\ \label{BHq}
   \q{3} |\Psi\rangle &= \frac{1}{\q{3}(E+i\varepsilon-\hat{H})\q{3}}
   \q{3} \hat{H} \qbar{3} |\Psi\rangle\;,
\end{align}
we can connect $\opty{3}$ to the zero-energy Bloch-Horowitz operator.
Specifically, if $|\Psi_3\rangle$ is a zero-energy 3-body 
eigenstate of $\hat{H}$, and if there is a 3-hole state 
$|\alpha_3\rangle$ such that
\ba\label{eq:alfpsi}
\opty{2}|\alpha_3\rangle&\approx\rob{\q{1}\p{2}+\q{2}\p{1}}|\Psi_3\rangle
\ea
then
\be\label{eq:t3psi}
   \opty{3}\ket{\alpha_3}\approx\q{3}\ket{\Psi_3}
\ee
and we can identify the matrix-elements of $\opty{3}$ with the $\q{3}$ 
components of the zero-energy solutions of the Schr{\"o}dinger equation.
In the next section we will argue that eq. \eqref{eq:t3psi} approximately holds.
%=============================================================================
\subsection{The 3-body zero-energy eigenstate}

We first note that a zero-energy 3-body eigenstate of the Schr{\"o}dinger equation, $\hat{H}\ket{\Psi_3}=0$, can be formally expanded in the CC method as
\ba
   |\Psi_3\rangle={\cal N}_3^{-1}e^{\opti{}}|\Phi_{0}\rangle =
   {\cal N}_3^{-1}(1+\opti{2}+\opti{3})|\tilde{\Phi}_{0}\rangle,
\ea
where $\opti{2},\opti{3}$ are the 3-body cluster operators, and 
\ba
   {\cal N}_3^2=
         1+\Tr\rob{\opti{2}^\dagger\opti{2}}+\Tr\rob{\opti{3}^\dagger\opti{3}}
\;
\ea
is a normalization factor.
Working with the momentum basis, we note that whereas the $A$-body operators $\opt{k}$ are defined with respect to the $A$-body
Fermi level $\kf$, the 3-body operators $\opti{2},\opti{3}$ are defined with respect to a
3-body reference state which we denote as $\ket{000}$, to indicate that it corresponds to single particle states with momentum
which is either zero or very close to zero.
We note that in this case the state $\ket{000}$ acts as a closed-shell state as the other possible Slater-determinants with zero momenta holes have different conserved quantum numbers, such as of $\hat{J}_z$,
and cannot contribute to $\ket{\Psi_3}$.
%For this reason $\opti{1}=0$ and we can understand that eq. \eqref{alfpsi} holds for $\kf\to0$ with $\ket{\alpha_3}$ being the Slater-determinant.
It follows that 
\ba\label{psi3}
|\Psi_3\rangle&={\cal N}_3^{-1}\left(\ket{000}
   +\frac{1}{2}\sum_{lm} \tilde{t}^{lm}_{00}\ket{lm0}+\frac{1}{2}\sum_{de} \tilde{t}^{de}_{00}\ket{de0}
   \right. \cr & \left.
   +\frac{1}{6}\sum_{lmn} \tilde{t}^{lmn}_{000}\ket{lmn}+\frac{1}{2}\sum_{dlm} \tilde{t}^{dlm}_{000}\ket{dlm}
   \right. \cr & \left.
   +\frac{1}{2}\sum_{del} \tilde{t}^{del}_{000}\ket{del}+\frac{1}{6}\sum_{def} \tilde{t}^{def}_{000}\ket{def}
   \right)\;.
\ea
Here, we keep notating the states according to the $A$-body Fermi level, e.g. $d,e,f$ correspond to particle states while $l,m,n$ to hole states.
As a result, terms such as $\ket{dl0}$ cannot appear in the expansion,
as momentum conservation implies that if the state $d$ is above the 
Fermi level so must be $l$.

Before substituting the 3-body wave-function \eqref{psi3} into the $Q$-space Bloch-Horowitz equation \eqref{BHq} we note that: 
\begin{enumerate*}[label=\itshape(\roman*)]
\item The operator $\qbar{3}$ kills the $3p0h$ states $\ket{def}$.
\item For 2-body interactions the term $\q{3} \hat{H} \qbar{3}$ annihilates the $0p3h$ states,
\end{enumerate*}
thus 
\ba\label{eq:qhq}
   \q{3} \hat{H} \qbar{3}|\Psi_3\rangle = &
       \q{3} \hat{H} \qbar{3}| \Psi_3^{(1p,2p)}\rangle
\ea
where
\ba\label{eq:1p2p}
  |\Psi_3^{(1p,2p)}\rangle & \equiv
  \rob{\q{1}\p{2}+\q{2}\p{1}}\ket{\Psi_3} = 
  \cr & =
   \f{{\cal N}_3^{-1}}{2}\!\left(
   \sum_{de} \tilde{t}^{de}_{00}\ket{de0}
      +\sum_{del} \tilde{t}^{del}_{000}\ket{del}
      \right. \cr & \hspace{5em} \left.
      +\sum_{dlm} \tilde{t}^{dlm}_{000}\ket{dlm}
      \right).
\ea

Inspecting Eq. \eqref{eq:1p2p} we note that the last term on the rhs is  zero unless $p_d<2\kf$. It follows that this term must vanish if we to consider a very dilute $A$-body system, i.e.  the limit $\kf\to 0$. Interestingly, in this limit also the first two terms coincide as $\tilde{t}^{del}_{000}\to \tilde{t}^{de0}_{000}$ with $l\to 0$.
Hence, in the limit $\kf\to0$, 
  $\q{3}\hat{H}\qbar{3}\ket{\Psi_3} \approx 
   2{\cal N}_3^{-1}\q{3}\hat{H}\opti{2}\ket{000}$.
Recalling now that Eq. \eqref{eq:t3meeq} is an asymptotic equation derived in the limit $p_a,p_b,p_c\to\infty$, and that in this limit 
$\opti{2}\to\opty{2}$, we may conclude that for $i,j,k=0$ the asymptotic
3-body cluster operator 
$\opty{3}$, Eq. \eqref{eq:t3inf}, can be redefined replacing $\opty{2}$ with $\opti{2}$. 
The resulting operator admits 
\ba
\opty{3}\ket{\alpha_3}\approx\q{3}\ket{\Psi_3}
\ea
with $\ket{\alpha_3}\equiv2{\cal N}_3^{-1}\ket{000}$.
% is thus exactly a Bloch-Horowitz operator and  $\opty{3}\ket{\Psi_3}=\q{3}\ket{\Psi_3}$.

Considering now the nuclear matter 
in the limit of dense matter (i.e. $\kf$ very large compared to the Fermi momentum at saturation density), we have 
%$\sum_{def}\tilde{t}^{def}_{000}\ket{def}\to\opty{3}\ket{000}$ and
$\f{1}{2}\sum_{de}\tilde{t}^{de}_{00}\ket{de0}\to \opty{2}\ket{000}$. Under this condition we also expect that the 
$2p1h$ terms $\f{1}{2}\sum_{del} \tilde{t}^{del}_{000}\ket{del}$  
are dominated by 2-body rather than 3-body correlations, i.e. most 
contributions will come from states with $p_l\ll p_d,p_e$. Therefore 
there is a 3-hole state $|\alpha_3^{(2p)}\rangle$ such that
$\f{1}{2}\sum_{del} \tilde{t}^{del}_{000}\ket{del}\approx \opty{2}|\alpha_3^{(2p)}\rangle$.
The $1p2h$ term $\f{1}{2}\sum_{dlm} \tilde{t}^{dlm}_{000}\ket{dlm}$ is clearly zero if the momentum of state $d$,
$p_d$, is larger than $2\kf$. We also expect that the main contribution of this term will appear when $p_d,p_l\approx \kf$ and the 3rd momentum is approximately zero.
Here again we can find a $0p3h$ state such that
$\f{1}{2}\sum_{dlm} \tilde{t}^{dlm}_{000}\ket{dlm}\approx \opty{2}|\alpha_3^{(1p)}\rangle$.
This observation implies that there is a $0p3h$ state $\alpha_3$
such that
\be
  |\Psi_3^{(1p,2p)}\rangle \approx \opty{2}|\alpha_3\rangle,
\ee
and hence also in this limit 
\be
  \opty{3}\ket{\alpha_3}\approx\q{3}\ket{\Psi_3}.
\ee
This relation holds also for any value of $\kf$ if we consider the most asymptotic high-momentum
contribution to $\opty{3}$,
hence we expect it to approximately hold for finite nuclei as well.

Summarizing this discussion we conclude that, as in the 2-body case,
\begin{enumerate*}[label=(\itshape\roman*)]
\item The asymptotic high-momentum behavior of $\opt{3}$ is related to a 3-body zero-energy eigen-function of the Schr{\"o}dinger equation.
\item This high momenta behavior is universal in the limited sense.
\end{enumerate*}
%=============================================================================
% Summary
% ============================================
\section{Summary}\label{sec:summary}
The CC method was utilized to set 
a more rigorous foundations for the successful GCF.
To this end we have computed the 2- and 3-body cluster operators in the high momentum limit and showed that they
act as the Bloch-Horowitz operators for the 2- and 3-body zero-energy eigen-states of the Schr{\" o}dinger equation.
We therefore concluded that the 2- and 3-body cluster operators in this high momentum limit are universal in the limited sense, that they  do not depend on the system but do depend on the potential.

The presented method is systematic and opens up the path for including higher order corrections to the GCF.  
A more complete discussion regarding the asymptotic wave function factorization is postponed to a  forthcoming article.
We note that our results may be useful for general CC computations, as   asymptotic expressions or approximations for the cluster operators.

% =============================================
% Acknowledgments
% ============================================
\begin{acknowledgments}
This research was supported by the ISRAEL SCIENCE FOUNDATION 
(grant No. 1086/21).
The work of S. Beck was also supported by the Israel Ministry of Science and Technology (MOST).
R. Weiss was supported by the Laboratory Directed Research and Development program of Los Alamos National Laboratory under project number 20210763PRD1.
\end{acknowledgments}
% =============================================
% APPENDIX
% ============================================
\appendix
%=============================================
% The asymptotics of $\opt{2}$
% ============================================
\section{The asymptotics of $\opt{2}$}\label{sec:t2asym}

To show that indeed the full 2-body amplitude $\opt{2}$ coincides with $\opty{2}$ in the limit
$p_a,p_b\to\infty$ we seek an iterative solution for Eq. \eqref{eq:fullteq2} \,\cite{shavit2009}.
To this end we denote by $\opt{2}^{(k)}$ the approximate solution of $\opt{2}$ after $k$ iterations. Taking the asymptotic solution to be our initial guess $\opt{2}^{(0)}=\opty{2}$, we can write 
\be
   \opt{2}^{(k)}=\opty{2}+\optd{(1)}_2+\optd{(2)}_2+\cdots+\optd{(k)}_2,
\ee
where the $k$th correction $\optd{(k)}_2$ is obtained by substituting $\opt{2}=\opt{2}^{(k-1)}+\optd{(k)}_2$ in \eqref{eq:fullteq2} and solving the linearized equation.

The equation for $\optd{(1)}_2$ reads 
\ba
   0 &= \bra{\Phi_{ij}^{ab}}
   \Vop + [\hat{H}_{0},\opty{2}]+[\hat{H}_{0},\optd{(1)}_2]+[\Vop,\opty{2}]
     \cr & 
     +[\Vop,\optd{(1)}_2]+\frac{1}{2}[[\Vop,\opty{2}],\opty{2}]+[\Vop,\opt{3}]+[\Vop,\opt{4}]
     \cr&
     +\f{1}{2}[[\Vop,\opty{2}],\optd{(1)}_2]+\f{1}{2}[[\Vop,\optd{(1)}_2],\opty{2}]\ket{\Phi_0}
\ea
Utilizing eq. \eqref{eq:t2univ} we obtain
\ba\label{eq:dt1_2}
   \Delta& t^{(1)ab}_{\quad ij} =
   -\frac{(\Vop\opty{2})_\text{res}}{E^{ab}_{ij}}
% +\f{E^{ij}}{E^{ab}E^{ab}}(\opty{2}\Vop+\Vop)_\text{res}
+\f{E^{ij}}{E^{ab}_{ij}}\rob{t^\infty}^{ab}_{ij}-\frac{1}{E^{ab}_{ij}}\bra{\Phi_{ij}^{ab}}
   \cr & \hspace{1em}
      [\Vop,\optd{(1)}_2]
   + \f{1}{2}[[\Vop,\opty{2}],\opty{2}]+[\Vop,\opt{3}]+[\Vop,\opt{4}]
   \cr & \hspace{1em}+\f{1}{2}[[\Vop,\opty{2}],\optd{(1)}_2]+\f{1}{2}[[\Vop,\optd{(1)}_2],\opty{2}]
   \ket{\Phi_0},
\ea
where $(\Vop \opty{2})_\text{res}$ stands for the terms that appear in \eqref{eq:t2me}
but are not included in \eqref{eq:t2univ}.
Asymptotically, as $p_a,p_b\to\infty$, the source terms should dominate
\begin{align}
  \Delta t^{(1)ab}_{\quad ij} \to& 
   -\frac{(\Vop\opty{2})_\text{res}}{E^{ab}_{ij}}
%    +\f{E^{ij}}{E^{ab}E^{ab}}(\opty{2}\Vop+\Vop)_\text{res}
+\f{E^{ij}}{E^{ab}_{ij}}\rob{t^\infty}^{ab}_{ij}-\frac{1}{E^{ab}_{ij}}\times
   \cr &\hspace{-3em}
   \bra{\Phi_{ij}^{ab}} \f{1}{2}[[\Vop,\opty{2}],\opty{2}]+[\Vop,\opt{3}]+[\Vop,\opt{4}]
   \ket{\Phi_0}.
\end{align}
Using the momentum arguments presented above and using the
% fact that the 3- and 4-body amplitudes behave asymptotically as $(\Vop \opty{2})_\text{res}$, it can be seen that all of these terms
inherent hierarchy, the 3- and 4-body terms and $(\Vop \opty{2})_\text{res}$ are suppressed by a factor $\langle\Vop\rangle/E^{ab}_{ij}$
(or $E^{ij}/E^{ab}_{ij}$)
compared to $\opty{2}$.
Hence asymptotically $\optd{(1)}_2\ll\opty{2}$.
By iterating the process one can see that asymptotically the higher order $k>1$ corrections 
are suppressed by a factor of the order $(\langle\Vop\rangle/E^{ab}_{ij})^k$.
This completes the iterative proof that $t^{ab}_{ij}\to\rob{t^\infty}^{ab}_{ij}$, and therefore in the high momentum limit we can replace the 2-body cluster operator $\opt{2}$ with the operator $\opty{2}$. 

%=============================================
% T2 as the bloch Horowitz operator
% ============================================
\section{$\opty{2}$ as the Bloch-Horowitz operator}\label{sec:t2_to_BH}
Recalling that $\Gop\rob{E}=\f{1}{E+i\varepsilon-\hat{H}_0}$ and that it commutes with the projection operators $\q{2},\p{2}$ we can write for the $\q{2}$ subspace
% \ba
%   \opty{2}&=\f{1}{1-\Gop\rob{E}\q{2}\Vop}\q{2}\Gop\rob{E}\hat{H}\p{2}
%   \cr&=
%   \f{1}{\q{2}-\Gop\rob{E}\q{2}\Vop}\q{2}\Gop\rob{E}\hat{H}\p{2}
%   \cr &=
%   \frac{1}{\Gop\rob{E}\q{2}(E+i\varepsilon-\hat{H}_0-\Vop)}
%   \Gop\rob{E}\q{2}\hat{H}\p{2}
%   \\&=
%   \f{1}{\q{2}(E+i\varepsilon-\hat{H})}\q{2}\hat{H}\p{2}
%   \;.
% \ea
\ba
  \opty{2}&=\f{1}{1-\q{2}\Gop\rob{E}\Vop}\q{2}\Gop\rob{E}\hat{H}\p{2}
  \cr&=
  \f{1}{\q{2}-\Gop\rob{E}\q{2}\Vop\q{2}}\Gop\rob{E}\q{2}\hat{H}\p{2}
  \cr &=
  \f{1}{\q{2}(E+i\varepsilon-\hat{H}_0-\Vop)\q{2}}\q{2}\hat{H}\p{2}
  \\&=
  \f{1}{\q{2}(E+i\varepsilon-\hat{H})\q{2}}\q{2}\hat{H}\p{2}
  \;.
\ea
Taking the value $E=0$ it can be rewritten as 
\ba
   \opty{2}=
   \f{1}{\q{2}(0-\hat{H})\q{2}}\q{2}\hat{H}\p{2}\;.
\ea

%=============================================
% The asymptotics of T3
% ============================================

\section{The asymptotics of $\opt{3}$}\label{sec:asymt3}
To show that $ t^{abc}_{ijk}\to\rob{t^\infty}_{ijk}^{abc}$ as $p_a,p_b,p_c\to\infty$
we substitute $\opt{3}=\opty{3}+\optd{}_3$ in eq. \eqref{eq:fullteq3} and solve for $\optd{}_3$ after using the definition of $\opty{3}$ in eq. \eqref{eq:t3inf}.
% Due to the inherent hierarchy the term $[[\Vop,\opt{2}],\opt{3}]$ and $\opt{4},\opt{5}$ can be neglected.
Moreover, as explained at section \ref{sec:cc amp high}, $\f{1}{2}[[\Vop,\opt{2}],\opt{2}]\ll[\Vop,\opt{2}]$ in the limit $p_a,p_b,p_c\to\infty$, hence the equation for $\Delta t^{abc}_{ijk}$ becomes
\ba
\Delta &t^{abc}_{ijk}=-\f{(\Vop\opty{3})_\text{res}}{E^{abc}_{ijk}}
+\f{E^{ijk}}{E^{abc}_{ijk}}\rob{t^\infty}^{abc}_{ijk}
\\&-\f{1}{E^{abc}_{ijk}}\bra{\Phi_{ijk}^{abc}}
  [\hat{H}_{0},\optd{}_3]
   + [\Vop,\optd{}_2]+[\Vop,\optd{}_3]
  \cr&+\f{1}{2}[[\Vop,\opt{2}],\optd{}_3]
  +[\Vop,\opt{4}]+[\Vop,\opt{5}]
\ket{\Phi_0}
\;\ea
where $(\Vop \opty{3})_\text{res}$ stands for the terms that appear in \eqref{eq:t3me}
but are not included in \eqref{eq:t3meeq} and $\optd{}_2=\opt{2}-\opty{2}$.
Asymptotically, the source terms should dominate, and thus
\ba
\Delta t^{abc}_{ijk}\to&-\f{(\Vop\opty{3})_\text{res}}{E^{abc}_{ijk}}
+\f{E^{ijk}}{E^{abc}_{ijk}}\rob{t^\infty}^{abc}_{ijk}-\f{1}{E^{abc}_{ijk}}\times
\\&\bra{\Phi_{ijk}^{abc}}[\Vop,\optd{}_2]+[\Vop,\opt{4}]+[\Vop,\opt{5}]\ket{\Phi_0}
\;.\ea
The terms in the first row are trivially much smaller than $(t^\infty)^{abc}_{ijk}$.
Also, the 4- and 5-body terms are also much smaller than $(t^\infty)^{abc}_{ijk}$
due to hierarchy and the suppression of 
% as they both behave as a 3-body amplitude and are suppressed by
the factor $\f{1}{E^{abc}_{ijk}}$.
 For the term $[\Vop,\optd{}_2]$ we can use the results of the previous section $\rob{\Delta t}^{cd}_{jk}\ll\rob{t^\infty}^{cd}_{jk}$ and note, from eq. \eqref{eq:t3inf}, that $\rob{t^\infty}^{abc}_{ijk}\sim -\f{1}{E^{abc}_{ijk}}{\cal \hat{S}}_{abc}\reb{{\cal \hat{S}}_{ijk}\reb{V^{ab}_{id}\rob{t^\infty}^{cd}_{jk}}}$. Altogether we get the desired reuslt $\rob{\Delta t}^{abc}_{ijk}\ll\rob{t^\infty}^{abc}_{ijk}$, i.e. $t^{abc}_{ijk}\to \rob{t^\infty}^{abc}_{ijk}$. 

\bibliography{references}

\begin{thebibliography}{59}%
\makeatletter
\providecommand \@ifxundefined [1]{%
 \@ifx{#1\undefined}
}%
\providecommand \@ifnum [1]{%
 \ifnum #1\expandafter \@firstoftwo
 \else \expandafter \@secondoftwo
 \fi
}%
\providecommand \@ifx [1]{%
 \ifx #1\expandafter \@firstoftwo
 \else \expandafter \@secondoftwo
 \fi
}%
\providecommand \natexlab [1]{#1}%
\providecommand \enquote  [1]{``#1''}%
\providecommand \bibnamefont  [1]{#1}%
\providecommand \bibfnamefont [1]{#1}%
\providecommand \citenamefont [1]{#1}%
\providecommand \href@noop [0]{\@secondoftwo}%
\providecommand \href [0]{\begingroup \@sanitize@url \@href}%
\providecommand \@href[1]{\@@startlink{#1}\@@href}%
\providecommand \@@href[1]{\endgroup#1\@@endlink}%
\providecommand \@sanitize@url [0]{\catcode `\\12\catcode `\$12\catcode
  `\&12\catcode `\#12\catcode `\^12\catcode `\_12\catcode `\%12\relax}%
\providecommand \@@startlink[1]{}%
\providecommand \@@endlink[0]{}%
\providecommand \url  [0]{\begingroup\@sanitize@url \@url }%
\providecommand \@url [1]{\endgroup\@href {#1}{\urlprefix }}%
\providecommand \urlprefix  [0]{URL }%
\providecommand \Eprint [0]{\href }%
\providecommand \doibase [0]{https://doi.org/}%
\providecommand \selectlanguage [0]{\@gobble}%
\providecommand \bibinfo  [0]{\@secondoftwo}%
\providecommand \bibfield  [0]{\@secondoftwo}%
\providecommand \translation [1]{[#1]}%
\providecommand \BibitemOpen [0]{}%
\providecommand \bibitemStop [0]{}%
\providecommand \bibitemNoStop [0]{.\EOS\space}%
\providecommand \EOS [0]{\spacefactor3000\relax}%
\providecommand \BibitemShut  [1]{\csname bibitem#1\endcsname}%
\let\auto@bib@innerbib\@empty
%</preamble>
\bibitem [{\citenamefont {Ciofi~degli Atti}(2015)}]{Atti:2015eda}%
  \BibitemOpen
  \bibfield  {author} {\bibinfo {author} {\bibfnamefont {C.}~\bibnamefont
  {Ciofi~degli Atti}},\ }\bibfield  {title} {\bibinfo {title} {{In-medium
  short-range dynamics of nucleons: Recent theoretical and experimental
  advances}},\ }\href {https://doi.org/10.1016/j.physrep.2015.06.002}
  {\bibfield  {journal} {\bibinfo  {journal} {Phys. Rept.}\ }\textbf {\bibinfo
  {volume} {590}},\ \bibinfo {pages} {1} (\bibinfo {year} {2015})}\BibitemShut
  {NoStop}%
%%CITATION = PRPLC,590,1;%%
\bibitem [{\citenamefont {Hen}\ \emph {et~al.}(2017)\citenamefont {Hen},
  \citenamefont {Miller}, \citenamefont {Piasetzky},\ and\ \citenamefont
  {Weinstein}}]{Hen:2016kwk}%
  \BibitemOpen
  \bibfield  {author} {\bibinfo {author} {\bibfnamefont {O.}~\bibnamefont
  {Hen}}, \bibinfo {author} {\bibfnamefont {G.~A.}\ \bibnamefont {Miller}},
  \bibinfo {author} {\bibfnamefont {E.}~\bibnamefont {Piasetzky}},\ and\
  \bibinfo {author} {\bibfnamefont {L.~B.}\ \bibnamefont {Weinstein}},\
  }\bibfield  {title} {\bibinfo {title} {{Nucleon-Nucleon Correlations,
  Short-lived Excitations, and the Quarks Within}},\ }\href
  {https://doi.org/10.1103/RevModPhys.89.045002} {\bibfield  {journal}
  {\bibinfo  {journal} {Rev. Mod. Phys.}\ }\textbf {\bibinfo {volume} {89}},\
  \bibinfo {pages} {045002} (\bibinfo {year} {2017})}\BibitemShut {NoStop}%
\bibitem [{\citenamefont {Frankfurt}\ and\ \citenamefont
  {Strikman}(1981)}]{Frankfurt81}%
  \BibitemOpen
  \bibfield  {author} {\bibinfo {author} {\bibfnamefont {L.~L.}\ \bibnamefont
  {Frankfurt}}\ and\ \bibinfo {author} {\bibfnamefont {M.~I.}\ \bibnamefont
  {Strikman}},\ }\bibfield  {title} {\bibinfo {title} {High-energy phenomena,
  short-range nuclear structure and qcd},\ }\href@noop {} {\bibfield  {journal}
  {\bibinfo  {journal} {Phys. Rep.}\ }\textbf {\bibinfo {volume} {76}},\
  \bibinfo {pages} {215} (\bibinfo {year} {1981})}\BibitemShut {NoStop}%
\bibitem [{\citenamefont {Frankfurt}\ and\ \citenamefont
  {Strikman}(1988)}]{Frankfurt88}%
  \BibitemOpen
  \bibfield  {author} {\bibinfo {author} {\bibfnamefont {L.}~\bibnamefont
  {Frankfurt}}\ and\ \bibinfo {author} {\bibfnamefont {M.}~\bibnamefont
  {Strikman}},\ }\bibfield  {title} {\bibinfo {title} {Hard nuclear processes
  and microscopic nuclear structure},\ }\href@noop {} {\bibfield  {journal}
  {\bibinfo  {journal} {Phys. Rep.}\ }\textbf {\bibinfo {volume} {160}},\
  \bibinfo {pages} {235 } (\bibinfo {year} {1988})}\BibitemShut {NoStop}%
\bibitem [{\citenamefont {Tang}\ \emph {et~al.}(2003)\citenamefont {Tang} \emph
  {et~al.}}]{tang03}%
  \BibitemOpen
  \bibfield  {author} {\bibinfo {author} {\bibfnamefont {A.}~\bibnamefont
  {Tang}} \emph {et~al.},\ }\bibfield  {title} {\bibinfo {title} {{n-p short
  range correlations from (p,2p + n) measurements}},\ }\href
  {https://doi.org/10.1103/PhysRevLett.90.042301} {\bibfield  {journal}
  {\bibinfo  {journal} {Phys. Rev. Lett.}\ }\textbf {\bibinfo {volume} {90}},\
  \bibinfo {pages} {042301} (\bibinfo {year} {2003})},\ \Eprint
  {https://arxiv.org/abs/nucl-ex/0206003} {arXiv:nucl-ex/0206003} \BibitemShut
  {NoStop}%
\bibitem [{\citenamefont {Piasetzky}\ \emph {et~al.}(2006)\citenamefont
  {Piasetzky}, \citenamefont {Sargsian}, \citenamefont {Frankfurt},
  \citenamefont {Strikman},\ and\ \citenamefont {Watson}}]{piasetzky06}%
  \BibitemOpen
  \bibfield  {author} {\bibinfo {author} {\bibfnamefont {E.}~\bibnamefont
  {Piasetzky}}, \bibinfo {author} {\bibfnamefont {M.}~\bibnamefont {Sargsian}},
  \bibinfo {author} {\bibfnamefont {L.}~\bibnamefont {Frankfurt}}, \bibinfo
  {author} {\bibfnamefont {M.}~\bibnamefont {Strikman}},\ and\ \bibinfo
  {author} {\bibfnamefont {J.~W.}\ \bibnamefont {Watson}},\ }\bibfield  {title}
  {\bibinfo {title} {Evidence for strong dominance of proton-neutron
  correlations in nuclei},\ }\href
  {https://doi.org/10.1103/PhysRevLett.97.162504} {\bibfield  {journal}
  {\bibinfo  {journal} {Phys. Rev. Lett.}\ }\textbf {\bibinfo {volume} {97}},\
  \bibinfo {pages} {162504} (\bibinfo {year} {2006})}\BibitemShut {NoStop}%
\bibitem [{\citenamefont {Shneor}\ \emph {et~al.}(2007)\citenamefont {Shneor}
  \emph {et~al.}}]{shneor07}%
  \BibitemOpen
  \bibfield  {author} {\bibinfo {author} {\bibfnamefont {R.}~\bibnamefont
  {Shneor}} \emph {et~al.} (\bibinfo {collaboration} {Jefferson Lab Hall A}),\
  }\bibfield  {title} {\bibinfo {title} {{Investigation of proton-proton
  short-range correlations via the C-12(e, e-prime pp) reaction}},\ }\href
  {https://doi.org/10.1103/PhysRevLett.99.072501} {\bibfield  {journal}
  {\bibinfo  {journal} {Phys. Rev. Lett.}\ }\textbf {\bibinfo {volume} {99}},\
  \bibinfo {pages} {072501} (\bibinfo {year} {2007})},\ \Eprint
  {https://arxiv.org/abs/nucl-ex/0703023} {arXiv:nucl-ex/0703023} \BibitemShut
  {NoStop}%
\bibitem [{\citenamefont {Subedi}\ \emph {et~al.}(2008)\citenamefont {Subedi}
  \emph {et~al.}}]{subedi08}%
  \BibitemOpen
  \bibfield  {author} {\bibinfo {author} {\bibfnamefont {R.}~\bibnamefont
  {Subedi}} \emph {et~al.},\ }\bibfield  {title} {\bibinfo {title} {{Probing
  Cold Dense Nuclear Matter}},\ }\href
  {https://doi.org/10.1126/science.1156675} {\bibfield  {journal} {\bibinfo
  {journal} {Science}\ }\textbf {\bibinfo {volume} {320}},\ \bibinfo {pages}
  {1476} (\bibinfo {year} {2008})},\ \Eprint {https://arxiv.org/abs/0908.1514}
  {arXiv:0908.1514 [nucl-ex]} \BibitemShut {NoStop}%
\bibitem [{\citenamefont {Korover}\ \emph {et~al.}(2014)\citenamefont
  {Korover}, \citenamefont {Muangma}, \citenamefont {Hen} \emph
  {et~al.}}]{korover14}%
  \BibitemOpen
  \bibfield  {author} {\bibinfo {author} {\bibfnamefont {I.}~\bibnamefont
  {Korover}}, \bibinfo {author} {\bibfnamefont {N.}~\bibnamefont {Muangma}},
  \bibinfo {author} {\bibfnamefont {O.}~\bibnamefont {Hen}}, \emph {et~al.},\
  }\bibfield  {title} {\bibinfo {title} {{Probing the Repulsive Core of the
  Nucleon-Nucleon Interaction via the 4He(e,e'pN) Triple-Coincidence
  Reaction}},\ }\href {https://doi.org/10.1103/PhysRevLett.113.022501}
  {\bibfield  {journal} {\bibinfo  {journal} {Phys. Rev. Lett.}\ }\textbf
  {\bibinfo {volume} {113}},\ \bibinfo {pages} {022501} (\bibinfo {year}
  {2014})}\BibitemShut {NoStop}%
\bibitem [{\citenamefont {Cohen}\ \emph {et~al.}(2018)\citenamefont {Cohen}
  \emph {et~al.}}]{Cohen:2018gzh}%
  \BibitemOpen
  \bibfield  {author} {\bibinfo {author} {\bibfnamefont {E.~O.}\ \bibnamefont
  {Cohen}} \emph {et~al.} (\bibinfo {collaboration} {CLAS Collaboration}),\
  }\bibfield  {title} {\bibinfo {title} {{Center of Mass Motion of Short-Range
  Correlated Nucleon Pairs studied via the $A(e,e'pp)$ Reaction}},\ }\href
  {https://doi.org/10.1103/PhysRevLett.121.092501} {\bibfield  {journal}
  {\bibinfo  {journal} {Phys. Rev. Lett.}\ }\textbf {\bibinfo {volume} {121}},\
  \bibinfo {pages} {092501} (\bibinfo {year} {2018})},\ \Eprint
  {https://arxiv.org/abs/1805.01981} {arXiv:1805.01981 [nucl-ex]} \BibitemShut
  {NoStop}%
%%CITATION = ARXIV:1805.01981;%%
\bibitem [{\citenamefont {Hen}\ \emph {et~al.}(2014)\citenamefont {Hen} \emph
  {et~al.}}]{hen14}%
  \BibitemOpen
  \bibfield  {author} {\bibinfo {author} {\bibfnamefont {O.}~\bibnamefont
  {Hen}} \emph {et~al.},\ }\bibfield  {title} {\bibinfo {title} {{Momentum
  sharing in imbalanced Fermi systems}},\ }\href
  {https://doi.org/10.1126/science.1256785} {\bibfield  {journal} {\bibinfo
  {journal} {Science}\ }\textbf {\bibinfo {volume} {346}},\ \bibinfo {pages}
  {614} (\bibinfo {year} {2014})},\ \Eprint {https://arxiv.org/abs/1412.0138}
  {arXiv:1412.0138 [nucl-ex]} \BibitemShut {NoStop}%
\bibitem [{\citenamefont {Duer}\ \emph {et~al.}(2019)\citenamefont {Duer} \emph
  {et~al.}}]{Duer:2018sxh}%
  \BibitemOpen
  \bibfield  {author} {\bibinfo {author} {\bibfnamefont {M.}~\bibnamefont
  {Duer}} \emph {et~al.} (\bibinfo {collaboration} {CLAS Collaboration}),\
  }\bibfield  {title} {\bibinfo {title} {{Direct Observation of Proton-Neutron
  Short-Range Correlation Dominance in Heavy Nuclei}},\ }\href
  {https://doi.org/10.1103/PhysRevLett.122.172502} {\bibfield  {journal}
  {\bibinfo  {journal} {Phys. Rev. Lett.}\ }\textbf {\bibinfo {volume} {122}},\
  \bibinfo {pages} {172502} (\bibinfo {year} {2019})},\ \Eprint
  {https://arxiv.org/abs/1810.05343} {arXiv:1810.05343 [nucl-ex]} \BibitemShut
  {NoStop}%
%%CITATION = ARXIV:1810.05343;%%
\bibitem [{\citenamefont {Schmidt}\ \emph {et~al.}(2020)\citenamefont {Schmidt}
  \emph {et~al.}}]{schmidt20}%
  \BibitemOpen
  \bibfield  {author} {\bibinfo {author} {\bibfnamefont {A.}~\bibnamefont
  {Schmidt}} \emph {et~al.} (\bibinfo {collaboration} {CLAS}),\ }\bibfield
  {title} {\bibinfo {title} {{Probing the core of the strong nuclear
  interaction}},\ }\href {https://doi.org/10.1038/s41586-020-2021-6} {\bibfield
   {journal} {\bibinfo  {journal} {Nature}\ }\textbf {\bibinfo {volume}
  {578}},\ \bibinfo {pages} {540} (\bibinfo {year} {2020})},\ \Eprint
  {https://arxiv.org/abs/2004.11221} {arXiv:2004.11221 [nucl-ex]} \BibitemShut
  {NoStop}%
\bibitem [{\citenamefont {Korover}\ \emph {et~al.}(2021)\citenamefont
  {Korover}, \citenamefont {Pybus}, \citenamefont {Schmidt}, \citenamefont
  {Hauenstein}, \citenamefont {Duer}, \citenamefont {Hen}, \citenamefont
  {Piasetzky}, \citenamefont {Weinstein}, \citenamefont {Higinbotham},
  \citenamefont {Adhikari}, \citenamefont {Adhikari}, \citenamefont {Amaryan},
  \citenamefont {Angelini}, \citenamefont {Atac}, \citenamefont {Barion},
  \citenamefont {Battaglieri}, \citenamefont {Beck}, \citenamefont
  {Bedlinskiy}, \citenamefont {Benmokhtar}, \citenamefont {Bianconi},
  \citenamefont {Biselli}, \citenamefont {Boiarinov}, \citenamefont {Briscoe},
  \citenamefont {Brooks}, \citenamefont {Bulumulla}, \citenamefont {Burkert},
  \citenamefont {Carman}, \citenamefont {Celentano}, \citenamefont {Chatagnon},
  \citenamefont {Chetry}, \citenamefont {Clark}, \citenamefont {Clary},
  \citenamefont {Cole}, \citenamefont {Contalbrigo}, \citenamefont {Crede},
  \citenamefont {Cruz-Torres}, \citenamefont {D'Angelo}, \citenamefont {{De
  Vita}}, \citenamefont {Defurne}, \citenamefont {Denniston}, \citenamefont
  {Deur}, \citenamefont {Diehl}, \citenamefont {Djalali}, \citenamefont
  {Dupre}, \citenamefont {Egiyan}, \citenamefont {Ehrhart}, \citenamefont {{El
  Alaoui}}, \citenamefont {{El Fassi}}, \citenamefont {Elouadrhiri},
  \citenamefont {Eugenio}, \citenamefont {Fersch}, \citenamefont {Filippi},
  \citenamefont {Forest}, \citenamefont {Gavalian}, \citenamefont {Girod},
  \citenamefont {Golovatch}, \citenamefont {Gothe}, \citenamefont {Griffioen},
  \citenamefont {Guidal}, \citenamefont {Hafidi}, \citenamefont {Hakobyan},
  \citenamefont {Harrison}, \citenamefont {Hattawy}, \citenamefont {Hayward},
  \citenamefont {Heddle}, \citenamefont {Hicks}, \citenamefont {Holtrop},
  \citenamefont {Ilieva}, \citenamefont {Ireland}, \citenamefont {Isupov},
  \citenamefont {Jenkins}, \citenamefont {Jo}, \citenamefont {Joo},
  \citenamefont {Joosten}, \citenamefont {Keller}, \citenamefont {Khachatryan},
  \citenamefont {Khanal}, \citenamefont {Khandaker}, \citenamefont {Kim},
  \citenamefont {Kim}, \citenamefont {Klein}, \citenamefont {Kubarovsky},
  \citenamefont {Lanza}, \citenamefont {Leali}, \citenamefont {Lenisa},
  \citenamefont {Livingston}, \citenamefont {Lucherini}, \citenamefont
  {MacGregor}, \citenamefont {Marchand}, \citenamefont {Markov}, \citenamefont
  {Marsicano}, \citenamefont {Mascagna}, \citenamefont {McKinnon},
  \citenamefont {{Mey-Tal Beck}}, \citenamefont {Mineeva}, \citenamefont
  {Mirazita}, \citenamefont {Movsisyan}, \citenamefont {{Munoz Camacho}},
  \citenamefont {Mustapha}, \citenamefont {Nadel-Turonski}, \citenamefont
  {Neupane}, \citenamefont {Niculescu}, \citenamefont {Osipenko}, \citenamefont
  {Ostrovidov}, \citenamefont {Paolone}, \citenamefont {Pappalardo},
  \citenamefont {Paremuzyan}, \citenamefont {Pasyuk}, \citenamefont {Phelps},
  \citenamefont {Pogorelko}, \citenamefont {Price}, \citenamefont {Prok},
  \citenamefont {Protopopescu}, \citenamefont {Raue}, \citenamefont {Ripani},
  \citenamefont {Ritman}, \citenamefont {Rizzo}, \citenamefont {Rosner},
  \citenamefont {Rowley}, \citenamefont {Sabatié}, \citenamefont {Salgado},
  \citenamefont {Schumacher}, \citenamefont {Segarra}, \citenamefont
  {Sharabian}, \citenamefont {Shrestha}, \citenamefont {Sokhan}, \citenamefont
  {Soto}, \citenamefont {Sparveris}, \citenamefont {Stepanyan}, \citenamefont
  {Strakovsky}, \citenamefont {Strauch}, \citenamefont {Tan}, \citenamefont
  {Tyler}, \citenamefont {Ungaro}, \citenamefont {Venturelli}, \citenamefont
  {Voskanyan}, \citenamefont {Voutier}, \citenamefont {Wang}, \citenamefont
  {Watts}, \citenamefont {Wei}, \citenamefont {Wood}, \citenamefont
  {Zachariou}, \citenamefont {Zhang}, \citenamefont {Zhao},\ and\ \citenamefont
  {Zheng}}]{Korover:2021}%
  \BibitemOpen
  \bibfield  {author} {\bibinfo {author} {\bibfnamefont {I.}~\bibnamefont
  {Korover}}, \bibinfo {author} {\bibfnamefont {J.}~\bibnamefont {Pybus}},
  \bibinfo {author} {\bibfnamefont {A.}~\bibnamefont {Schmidt}}, \bibinfo
  {author} {\bibfnamefont {F.}~\bibnamefont {Hauenstein}}, \bibinfo {author}
  {\bibfnamefont {M.}~\bibnamefont {Duer}}, \bibinfo {author} {\bibfnamefont
  {O.}~\bibnamefont {Hen}}, \bibinfo {author} {\bibfnamefont {E.}~\bibnamefont
  {Piasetzky}}, \bibinfo {author} {\bibfnamefont {L.}~\bibnamefont
  {Weinstein}}, \bibinfo {author} {\bibfnamefont {D.}~\bibnamefont
  {Higinbotham}}, \bibinfo {author} {\bibfnamefont {S.}~\bibnamefont
  {Adhikari}}, \bibinfo {author} {\bibfnamefont {K.}~\bibnamefont {Adhikari}},
  \bibinfo {author} {\bibfnamefont {M.}~\bibnamefont {Amaryan}}, \bibinfo
  {author} {\bibfnamefont {G.}~\bibnamefont {Angelini}}, \bibinfo {author}
  {\bibfnamefont {H.}~\bibnamefont {Atac}}, \bibinfo {author} {\bibfnamefont
  {L.}~\bibnamefont {Barion}}, \bibinfo {author} {\bibfnamefont
  {M.}~\bibnamefont {Battaglieri}}, \bibinfo {author} {\bibfnamefont
  {A.}~\bibnamefont {Beck}}, \bibinfo {author} {\bibfnamefont {I.}~\bibnamefont
  {Bedlinskiy}}, \bibinfo {author} {\bibfnamefont {F.}~\bibnamefont
  {Benmokhtar}}, \bibinfo {author} {\bibfnamefont {A.}~\bibnamefont
  {Bianconi}}, \bibinfo {author} {\bibfnamefont {A.}~\bibnamefont {Biselli}},
  \bibinfo {author} {\bibfnamefont {S.}~\bibnamefont {Boiarinov}}, \bibinfo
  {author} {\bibfnamefont {W.}~\bibnamefont {Briscoe}}, \bibinfo {author}
  {\bibfnamefont {W.}~\bibnamefont {Brooks}}, \bibinfo {author} {\bibfnamefont
  {D.}~\bibnamefont {Bulumulla}}, \bibinfo {author} {\bibfnamefont
  {V.}~\bibnamefont {Burkert}}, \bibinfo {author} {\bibfnamefont
  {D.}~\bibnamefont {Carman}}, \bibinfo {author} {\bibfnamefont
  {A.}~\bibnamefont {Celentano}}, \bibinfo {author} {\bibfnamefont
  {P.}~\bibnamefont {Chatagnon}}, \bibinfo {author} {\bibfnamefont
  {T.}~\bibnamefont {Chetry}}, \bibinfo {author} {\bibfnamefont
  {L.}~\bibnamefont {Clark}}, \bibinfo {author} {\bibfnamefont
  {B.}~\bibnamefont {Clary}}, \bibinfo {author} {\bibfnamefont
  {P.}~\bibnamefont {Cole}}, \bibinfo {author} {\bibfnamefont {M.}~\bibnamefont
  {Contalbrigo}}, \bibinfo {author} {\bibfnamefont {V.}~\bibnamefont {Crede}},
  \bibinfo {author} {\bibfnamefont {R.}~\bibnamefont {Cruz-Torres}}, \bibinfo
  {author} {\bibfnamefont {A.}~\bibnamefont {D'Angelo}}, \bibinfo {author}
  {\bibfnamefont {R.}~\bibnamefont {{De Vita}}}, \bibinfo {author}
  {\bibfnamefont {M.}~\bibnamefont {Defurne}}, \bibinfo {author} {\bibfnamefont
  {A.}~\bibnamefont {Denniston}}, \bibinfo {author} {\bibfnamefont
  {A.}~\bibnamefont {Deur}}, \bibinfo {author} {\bibfnamefont {S.}~\bibnamefont
  {Diehl}}, \bibinfo {author} {\bibfnamefont {C.}~\bibnamefont {Djalali}},
  \bibinfo {author} {\bibfnamefont {R.}~\bibnamefont {Dupre}}, \bibinfo
  {author} {\bibfnamefont {H.}~\bibnamefont {Egiyan}}, \bibinfo {author}
  {\bibfnamefont {M.}~\bibnamefont {Ehrhart}}, \bibinfo {author} {\bibfnamefont
  {A.}~\bibnamefont {{El Alaoui}}}, \bibinfo {author} {\bibfnamefont
  {L.}~\bibnamefont {{El Fassi}}}, \bibinfo {author} {\bibfnamefont
  {L.}~\bibnamefont {Elouadrhiri}}, \bibinfo {author} {\bibfnamefont
  {P.}~\bibnamefont {Eugenio}}, \bibinfo {author} {\bibfnamefont
  {R.}~\bibnamefont {Fersch}}, \bibinfo {author} {\bibfnamefont
  {A.}~\bibnamefont {Filippi}}, \bibinfo {author} {\bibfnamefont
  {T.}~\bibnamefont {Forest}}, \bibinfo {author} {\bibfnamefont
  {G.}~\bibnamefont {Gavalian}}, \bibinfo {author} {\bibfnamefont
  {F.}~\bibnamefont {Girod}}, \bibinfo {author} {\bibfnamefont
  {E.}~\bibnamefont {Golovatch}}, \bibinfo {author} {\bibfnamefont
  {R.}~\bibnamefont {Gothe}}, \bibinfo {author} {\bibfnamefont
  {K.}~\bibnamefont {Griffioen}}, \bibinfo {author} {\bibfnamefont
  {M.}~\bibnamefont {Guidal}}, \bibinfo {author} {\bibfnamefont
  {K.}~\bibnamefont {Hafidi}}, \bibinfo {author} {\bibfnamefont
  {H.}~\bibnamefont {Hakobyan}}, \bibinfo {author} {\bibfnamefont
  {N.}~\bibnamefont {Harrison}}, \bibinfo {author} {\bibfnamefont
  {M.}~\bibnamefont {Hattawy}}, \bibinfo {author} {\bibfnamefont
  {T.}~\bibnamefont {Hayward}}, \bibinfo {author} {\bibfnamefont
  {D.}~\bibnamefont {Heddle}}, \bibinfo {author} {\bibfnamefont
  {K.}~\bibnamefont {Hicks}}, \bibinfo {author} {\bibfnamefont
  {M.}~\bibnamefont {Holtrop}}, \bibinfo {author} {\bibfnamefont
  {Y.}~\bibnamefont {Ilieva}}, \bibinfo {author} {\bibfnamefont
  {D.}~\bibnamefont {Ireland}}, \bibinfo {author} {\bibfnamefont
  {E.}~\bibnamefont {Isupov}}, \bibinfo {author} {\bibfnamefont
  {D.}~\bibnamefont {Jenkins}}, \bibinfo {author} {\bibfnamefont
  {H.}~\bibnamefont {Jo}}, \bibinfo {author} {\bibfnamefont {K.}~\bibnamefont
  {Joo}}, \bibinfo {author} {\bibfnamefont {S.}~\bibnamefont {Joosten}},
  \bibinfo {author} {\bibfnamefont {D.}~\bibnamefont {Keller}}, \bibinfo
  {author} {\bibfnamefont {M.}~\bibnamefont {Khachatryan}}, \bibinfo {author}
  {\bibfnamefont {A.}~\bibnamefont {Khanal}}, \bibinfo {author} {\bibfnamefont
  {M.}~\bibnamefont {Khandaker}}, \bibinfo {author} {\bibfnamefont
  {A.}~\bibnamefont {Kim}}, \bibinfo {author} {\bibfnamefont {C.}~\bibnamefont
  {Kim}}, \bibinfo {author} {\bibfnamefont {F.}~\bibnamefont {Klein}}, \bibinfo
  {author} {\bibfnamefont {V.}~\bibnamefont {Kubarovsky}}, \bibinfo {author}
  {\bibfnamefont {L.}~\bibnamefont {Lanza}}, \bibinfo {author} {\bibfnamefont
  {M.}~\bibnamefont {Leali}}, \bibinfo {author} {\bibfnamefont
  {P.}~\bibnamefont {Lenisa}}, \bibinfo {author} {\bibfnamefont
  {K.}~\bibnamefont {Livingston}}, \bibinfo {author} {\bibfnamefont
  {V.}~\bibnamefont {Lucherini}}, \bibinfo {author} {\bibfnamefont
  {I.}~\bibnamefont {MacGregor}}, \bibinfo {author} {\bibfnamefont
  {D.}~\bibnamefont {Marchand}}, \bibinfo {author} {\bibfnamefont
  {N.}~\bibnamefont {Markov}}, \bibinfo {author} {\bibfnamefont
  {L.}~\bibnamefont {Marsicano}}, \bibinfo {author} {\bibfnamefont
  {V.}~\bibnamefont {Mascagna}}, \bibinfo {author} {\bibfnamefont
  {B.}~\bibnamefont {McKinnon}}, \bibinfo {author} {\bibfnamefont
  {S.}~\bibnamefont {{Mey-Tal Beck}}}, \bibinfo {author} {\bibfnamefont
  {T.}~\bibnamefont {Mineeva}}, \bibinfo {author} {\bibfnamefont
  {M.}~\bibnamefont {Mirazita}}, \bibinfo {author} {\bibfnamefont
  {A.}~\bibnamefont {Movsisyan}}, \bibinfo {author} {\bibfnamefont
  {C.}~\bibnamefont {{Munoz Camacho}}}, \bibinfo {author} {\bibfnamefont
  {B.}~\bibnamefont {Mustapha}}, \bibinfo {author} {\bibfnamefont
  {P.}~\bibnamefont {Nadel-Turonski}}, \bibinfo {author} {\bibfnamefont
  {K.}~\bibnamefont {Neupane}}, \bibinfo {author} {\bibfnamefont
  {G.}~\bibnamefont {Niculescu}}, \bibinfo {author} {\bibfnamefont
  {M.}~\bibnamefont {Osipenko}}, \bibinfo {author} {\bibfnamefont
  {A.}~\bibnamefont {Ostrovidov}}, \bibinfo {author} {\bibfnamefont
  {M.}~\bibnamefont {Paolone}}, \bibinfo {author} {\bibfnamefont
  {L.}~\bibnamefont {Pappalardo}}, \bibinfo {author} {\bibfnamefont
  {R.}~\bibnamefont {Paremuzyan}}, \bibinfo {author} {\bibfnamefont
  {E.}~\bibnamefont {Pasyuk}}, \bibinfo {author} {\bibfnamefont
  {W.}~\bibnamefont {Phelps}}, \bibinfo {author} {\bibfnamefont
  {O.}~\bibnamefont {Pogorelko}}, \bibinfo {author} {\bibfnamefont
  {J.}~\bibnamefont {Price}}, \bibinfo {author} {\bibfnamefont
  {Y.}~\bibnamefont {Prok}}, \bibinfo {author} {\bibfnamefont {D.}~\bibnamefont
  {Protopopescu}}, \bibinfo {author} {\bibfnamefont {B.}~\bibnamefont {Raue}},
  \bibinfo {author} {\bibfnamefont {M.}~\bibnamefont {Ripani}}, \bibinfo
  {author} {\bibfnamefont {J.}~\bibnamefont {Ritman}}, \bibinfo {author}
  {\bibfnamefont {A.}~\bibnamefont {Rizzo}}, \bibinfo {author} {\bibfnamefont
  {G.}~\bibnamefont {Rosner}}, \bibinfo {author} {\bibfnamefont
  {J.}~\bibnamefont {Rowley}}, \bibinfo {author} {\bibfnamefont
  {F.}~\bibnamefont {Sabatié}}, \bibinfo {author} {\bibfnamefont
  {C.}~\bibnamefont {Salgado}}, \bibinfo {author} {\bibfnamefont
  {R.}~\bibnamefont {Schumacher}}, \bibinfo {author} {\bibfnamefont
  {E.}~\bibnamefont {Segarra}}, \bibinfo {author} {\bibfnamefont
  {Y.}~\bibnamefont {Sharabian}}, \bibinfo {author} {\bibfnamefont
  {U.}~\bibnamefont {Shrestha}}, \bibinfo {author} {\bibfnamefont
  {D.}~\bibnamefont {Sokhan}}, \bibinfo {author} {\bibfnamefont
  {O.}~\bibnamefont {Soto}}, \bibinfo {author} {\bibfnamefont {N.}~\bibnamefont
  {Sparveris}}, \bibinfo {author} {\bibfnamefont {S.}~\bibnamefont
  {Stepanyan}}, \bibinfo {author} {\bibfnamefont {I.}~\bibnamefont
  {Strakovsky}}, \bibinfo {author} {\bibfnamefont {S.}~\bibnamefont {Strauch}},
  \bibinfo {author} {\bibfnamefont {J.}~\bibnamefont {Tan}}, \bibinfo {author}
  {\bibfnamefont {N.}~\bibnamefont {Tyler}}, \bibinfo {author} {\bibfnamefont
  {M.}~\bibnamefont {Ungaro}}, \bibinfo {author} {\bibfnamefont
  {L.}~\bibnamefont {Venturelli}}, \bibinfo {author} {\bibfnamefont
  {H.}~\bibnamefont {Voskanyan}}, \bibinfo {author} {\bibfnamefont
  {E.}~\bibnamefont {Voutier}}, \bibinfo {author} {\bibfnamefont
  {T.}~\bibnamefont {Wang}}, \bibinfo {author} {\bibfnamefont {D.}~\bibnamefont
  {Watts}}, \bibinfo {author} {\bibfnamefont {X.}~\bibnamefont {Wei}}, \bibinfo
  {author} {\bibfnamefont {M.}~\bibnamefont {Wood}}, \bibinfo {author}
  {\bibfnamefont {N.}~\bibnamefont {Zachariou}}, \bibinfo {author}
  {\bibfnamefont {J.}~\bibnamefont {Zhang}}, \bibinfo {author} {\bibfnamefont
  {Z.}~\bibnamefont {Zhao}},\ and\ \bibinfo {author} {\bibfnamefont
  {X.}~\bibnamefont {Zheng}},\ }\bibfield  {title} {\bibinfo {title}
  {12c(e,e'pn) measurements of short range correlations in the tensor-to-scalar
  interaction transition region},\ }\href
  {https://doi.org/https://doi.org/10.1016/j.physletb.2021.136523} {\bibfield
  {journal} {\bibinfo  {journal} {Physics Letters B}\ }\textbf {\bibinfo
  {volume} {820}},\ \bibinfo {pages} {136523} (\bibinfo {year}
  {2021})}\BibitemShut {NoStop}%
\bibitem [{\citenamefont {Frankfurt}\ \emph {et~al.}(1993)\citenamefont
  {Frankfurt}, \citenamefont {Strikman}, \citenamefont {Day},\ and\
  \citenamefont {Sargsyan}}]{frankfurt93}%
  \BibitemOpen
  \bibfield  {author} {\bibinfo {author} {\bibfnamefont {L.}~\bibnamefont
  {Frankfurt}}, \bibinfo {author} {\bibfnamefont {M.}~\bibnamefont {Strikman}},
  \bibinfo {author} {\bibfnamefont {D.}~\bibnamefont {Day}},\ and\ \bibinfo
  {author} {\bibfnamefont {M.}~\bibnamefont {Sargsyan}},\ }\bibfield  {title}
  {\bibinfo {title} {Evidence for short-range correlations from high q2 (e,e')
  reactions},\ }\href@noop {} {\bibfield  {journal} {\bibinfo  {journal} {Phys.
  Rev. C}\ }\textbf {\bibinfo {volume} {48}},\ \bibinfo {pages} {2451}
  (\bibinfo {year} {1993})}\BibitemShut {NoStop}%
\bibitem [{\citenamefont {Egiyan}\ \emph {et~al.}(2003)\citenamefont {Egiyan}
  \emph {et~al.}}]{egiyan02}%
  \BibitemOpen
  \bibfield  {author} {\bibinfo {author} {\bibfnamefont {K.}~\bibnamefont
  {Egiyan}} \emph {et~al.} (\bibinfo {collaboration} {CLAS Collaboration}),\
  }\href@noop {} {\bibfield  {journal} {\bibinfo  {journal} {Phys. Rev. C}\
  }\textbf {\bibinfo {volume} {68}},\ \bibinfo {pages} {014313} (\bibinfo
  {year} {2003})}\BibitemShut {NoStop}%
\bibitem [{\citenamefont {Egiyan}\ \emph {et~al.}(2006)\citenamefont {Egiyan}
  \emph {et~al.}}]{egiyan06}%
  \BibitemOpen
  \bibfield  {author} {\bibinfo {author} {\bibfnamefont {K.}~\bibnamefont
  {Egiyan}} \emph {et~al.} (\bibinfo {collaboration} {CLAS Collaboration}),\
  }\bibfield  {title} {\bibinfo {title} {Measurement of 2- and 3-nucleon short
  range correlation probabilities in nuclei},\ }\href@noop {} {\bibfield
  {journal} {\bibinfo  {journal} {Phys. Rev. Lett.}\ }\textbf {\bibinfo
  {volume} {96}},\ \bibinfo {pages} {082501} (\bibinfo {year}
  {2006})}\BibitemShut {NoStop}%
\bibitem [{\citenamefont {Fomin}\ \emph {et~al.}(2012)\citenamefont {Fomin}
  \emph {et~al.}}]{fomin12}%
  \BibitemOpen
  \bibfield  {author} {\bibinfo {author} {\bibfnamefont {N.}~\bibnamefont
  {Fomin}} \emph {et~al.},\ }\bibfield  {title} {\bibinfo {title} {New
  measurements of high-momentum nucleons and short-range structures in
  nuclei},\ }\href@noop {} {\bibfield  {journal} {\bibinfo  {journal} {Phys.
  Rev. Lett.}\ }\textbf {\bibinfo {volume} {108}},\ \bibinfo {pages} {092502}
  (\bibinfo {year} {2012})}\BibitemShut {NoStop}%
\bibitem [{\citenamefont {Schmookler}\ \emph {et~al.}(2019)\citenamefont
  {Schmookler} \emph {et~al.}}]{Schmookler:2019nvf}%
  \BibitemOpen
  \bibfield  {author} {\bibinfo {author} {\bibfnamefont {B.}~\bibnamefont
  {Schmookler}} \emph {et~al.} (\bibinfo {collaboration} {CLAS
  Collaboration}),\ }\bibfield  {title} {\bibinfo {title} {{Modified structure
  of protons and neutrons in correlated pairs}},\ }\href
  {https://doi.org/10.1038/s41586-019-0925-9} {\bibfield  {journal} {\bibinfo
  {journal} {Nature}\ }\textbf {\bibinfo {volume} {566}},\ \bibinfo {pages}
  {354} (\bibinfo {year} {2019})}\BibitemShut {NoStop}%
%%CITATION = NATUA,566,354;%%
\bibitem [{\citenamefont {Feldmeier}\ \emph {et~al.}(2011)\citenamefont
  {Feldmeier}, \citenamefont {Horiuchi}, \citenamefont {Neff},\ and\
  \citenamefont {Suzuki}}]{Feldmeier:2011qy}%
  \BibitemOpen
  \bibfield  {author} {\bibinfo {author} {\bibfnamefont {H.}~\bibnamefont
  {Feldmeier}}, \bibinfo {author} {\bibfnamefont {W.}~\bibnamefont {Horiuchi}},
  \bibinfo {author} {\bibfnamefont {T.}~\bibnamefont {Neff}},\ and\ \bibinfo
  {author} {\bibfnamefont {Y.}~\bibnamefont {Suzuki}},\ }\bibfield  {title}
  {\bibinfo {title} {{Universality of short-range nucleon-nucleon
  correlations}},\ }\href {https://doi.org/10.1103/PhysRevC.84.054003}
  {\bibfield  {journal} {\bibinfo  {journal} {Phys. Rev. C}\ }\textbf {\bibinfo
  {volume} {84}},\ \bibinfo {pages} {054003} (\bibinfo {year} {2011})},\
  \Eprint {https://arxiv.org/abs/1107.4956} {arXiv:1107.4956 [nucl-th]}
  \BibitemShut {NoStop}%
\bibitem [{\citenamefont {Alvioli}\ \emph
  {et~al.}(2013{\natexlab{a}})\citenamefont {Alvioli}, \citenamefont
  {Ciofi~degli Atti}, \citenamefont {Kaptari}, \citenamefont {Mezzetti},\ and\
  \citenamefont {Morita}}]{Alvioli:2012qa}%
  \BibitemOpen
  \bibfield  {author} {\bibinfo {author} {\bibfnamefont {M.}~\bibnamefont
  {Alvioli}}, \bibinfo {author} {\bibfnamefont {C.}~\bibnamefont {Ciofi~degli
  Atti}}, \bibinfo {author} {\bibfnamefont {L.~P.}\ \bibnamefont {Kaptari}},
  \bibinfo {author} {\bibfnamefont {C.~B.}\ \bibnamefont {Mezzetti}},\ and\
  \bibinfo {author} {\bibfnamefont {H.}~\bibnamefont {Morita}},\ }\bibfield
  {title} {\bibinfo {title} {{Nucleon momentum distributions, their
  spin-isospin dependence and short-range correlations}},\ }\href
  {https://doi.org/10.1103/PhysRevC.87.034603} {\bibfield  {journal} {\bibinfo
  {journal} {Phys. Rev.}\ }\textbf {\bibinfo {volume} {C87}},\ \bibinfo {pages}
  {034603} (\bibinfo {year} {2013}{\natexlab{a}})},\ \Eprint
  {https://arxiv.org/abs/1211.0134} {arXiv:1211.0134 [nucl-th]} \BibitemShut
  {NoStop}%
%%CITATION = ARXIV:1211.0134;%%
\bibitem [{\citenamefont {Wiringa}\ \emph {et~al.}(2014)\citenamefont
  {Wiringa}, \citenamefont {Schiavilla}, \citenamefont {Pieper},\ and\
  \citenamefont {Carlson}}]{wiringa14}%
  \BibitemOpen
  \bibfield  {author} {\bibinfo {author} {\bibfnamefont {R.~B.}\ \bibnamefont
  {Wiringa}}, \bibinfo {author} {\bibfnamefont {R.}~\bibnamefont {Schiavilla}},
  \bibinfo {author} {\bibfnamefont {S.~C.}\ \bibnamefont {Pieper}},\ and\
  \bibinfo {author} {\bibfnamefont {J.}~\bibnamefont {Carlson}},\ }\bibfield
  {title} {\bibinfo {title} {Nucleon and nucleon-pair momentum distributions in
  $a\le 12$},\ }\href@noop {} {\bibfield  {journal} {\bibinfo  {journal} {Phys.
  Rev. C}\ }\textbf {\bibinfo {volume} {89}},\ \bibinfo {pages} {024305}
  (\bibinfo {year} {2014})}\BibitemShut {NoStop}%
\bibitem [{\citenamefont {Rios}\ \emph {et~al.}(2014)\citenamefont {Rios},
  \citenamefont {Polls},\ and\ \citenamefont {Dickhoff}}]{Rios:2013zqa}%
  \BibitemOpen
  \bibfield  {author} {\bibinfo {author} {\bibfnamefont {A.}~\bibnamefont
  {Rios}}, \bibinfo {author} {\bibfnamefont {A.}~\bibnamefont {Polls}},\ and\
  \bibinfo {author} {\bibfnamefont {W.~H.}\ \bibnamefont {Dickhoff}},\
  }\bibfield  {title} {\bibinfo {title} {{Density and isospin asymmetry
  dependence of high-momentum components}},\ }\href
  {https://doi.org/10.1103/PhysRevC.89.044303} {\bibfield  {journal} {\bibinfo
  {journal} {Phys. Rev.}\ }\textbf {\bibinfo {volume} {C89}},\ \bibinfo {pages}
  {044303} (\bibinfo {year} {2014})},\ \Eprint
  {https://arxiv.org/abs/1312.7307} {arXiv:1312.7307 [nucl-th]} \BibitemShut
  {NoStop}%
%%CITATION = ARXIV:1312.7307;%%
\bibitem [{\citenamefont {Ryckebusch}\ \emph {et~al.}(2019)\citenamefont
  {Ryckebusch}, \citenamefont {Cosyn}, \citenamefont {Vieijra},\ and\
  \citenamefont {Casert}}]{Ryckebusch:2019oya}%
  \BibitemOpen
  \bibfield  {author} {\bibinfo {author} {\bibfnamefont {J.}~\bibnamefont
  {Ryckebusch}}, \bibinfo {author} {\bibfnamefont {W.}~\bibnamefont {Cosyn}},
  \bibinfo {author} {\bibfnamefont {T.}~\bibnamefont {Vieijra}},\ and\ \bibinfo
  {author} {\bibfnamefont {C.}~\bibnamefont {Casert}},\ }\bibfield  {title}
  {\bibinfo {title} {{Isospin composition of the high-momentum fluctuations in
  nuclei from asymptotic momentum distributions}},\ }\href
  {https://doi.org/10.1103/PhysRevC.100.054620} {\bibfield  {journal} {\bibinfo
   {journal} {Phys. Rev. C}\ }\textbf {\bibinfo {volume} {100}},\ \bibinfo
  {pages} {054620} (\bibinfo {year} {2019})},\ \Eprint
  {https://arxiv.org/abs/1907.07259} {arXiv:1907.07259 [nucl-th]} \BibitemShut
  {NoStop}%
\bibitem [{\citenamefont {Alvioli}\ \emph
  {et~al.}(2013{\natexlab{b}})\citenamefont {Alvioli}, \citenamefont {Ciofi
  Degli~Atti}, \citenamefont {Kaptari}, \citenamefont {Mezzetti},\ and\
  \citenamefont {Morita}}]{Alvioli:2013qyz}%
  \BibitemOpen
  \bibfield  {author} {\bibinfo {author} {\bibfnamefont {M.}~\bibnamefont
  {Alvioli}}, \bibinfo {author} {\bibfnamefont {C.}~\bibnamefont {Ciofi
  Degli~Atti}}, \bibinfo {author} {\bibfnamefont {L.~P.}\ \bibnamefont
  {Kaptari}}, \bibinfo {author} {\bibfnamefont {C.~B.}\ \bibnamefont
  {Mezzetti}},\ and\ \bibinfo {author} {\bibfnamefont {H.}~\bibnamefont
  {Morita}},\ }\bibfield  {title} {\bibinfo {title} {{Universality of
  nucleon-nucleon short-range correlations and nucleon momentum
  distributions}},\ }\href {https://doi.org/10.1142/S021830131330021X}
  {\bibfield  {journal} {\bibinfo  {journal} {Int. J. Mod. Phys.}\ }\textbf
  {\bibinfo {volume} {E22}},\ \bibinfo {pages} {1330021} (\bibinfo {year}
  {2013}{\natexlab{b}})},\ \Eprint {https://arxiv.org/abs/1306.6235}
  {arXiv:1306.6235 [nucl-th]} \BibitemShut {NoStop}%
%%CITATION = ARXIV:1306.6235;%%
\bibitem [{\citenamefont {Neff}\ \emph {et~al.}(2015)\citenamefont {Neff},
  \citenamefont {Feldmeier},\ and\ \citenamefont {Horiuchi}}]{neff15}%
  \BibitemOpen
  \bibfield  {author} {\bibinfo {author} {\bibfnamefont {T.}~\bibnamefont
  {Neff}}, \bibinfo {author} {\bibfnamefont {H.}~\bibnamefont {Feldmeier}},\
  and\ \bibinfo {author} {\bibfnamefont {W.}~\bibnamefont {Horiuchi}},\
  }\bibfield  {title} {\bibinfo {title} {Short-range correlations in nuclei
  with similarity renormalization group transformations},\ }\href
  {https://doi.org/10.1103/PhysRevC.92.024003} {\bibfield  {journal} {\bibinfo
  {journal} {Phys. Rev. C}\ }\textbf {\bibinfo {volume} {92}},\ \bibinfo
  {pages} {024003} (\bibinfo {year} {2015})}\BibitemShut {NoStop}%
\bibitem [{\citenamefont {Ryckebusch}\ \emph {et~al.}(2015)\citenamefont
  {Ryckebusch}, \citenamefont {Vanhalst},\ and\ \citenamefont
  {Cosyn}}]{ryckebusch15}%
  \BibitemOpen
  \bibfield  {author} {\bibinfo {author} {\bibfnamefont {J.}~\bibnamefont
  {Ryckebusch}}, \bibinfo {author} {\bibfnamefont {M.}~\bibnamefont
  {Vanhalst}},\ and\ \bibinfo {author} {\bibfnamefont {W.}~\bibnamefont
  {Cosyn}},\ }\bibfield  {title} {\bibinfo {title} {Stylized features of
  single-nucleon momentum distributions},\ }\href@noop {} {\bibfield  {journal}
  {\bibinfo  {journal} {Journal of Physics G: Nuclear and Particle Physics}\
  }\textbf {\bibinfo {volume} {42}},\ \bibinfo {pages} {055104} (\bibinfo
  {year} {2015})}\BibitemShut {NoStop}%
\bibitem [{\citenamefont {Alvioli}\ \emph {et~al.}(2008)\citenamefont
  {Alvioli}, \citenamefont {Ciofi~degli Atti},\ and\ \citenamefont
  {Morita}}]{Alvioli:2007zz}%
  \BibitemOpen
  \bibfield  {author} {\bibinfo {author} {\bibfnamefont {M.}~\bibnamefont
  {Alvioli}}, \bibinfo {author} {\bibfnamefont {C.}~\bibnamefont {Ciofi~degli
  Atti}},\ and\ \bibinfo {author} {\bibfnamefont {H.}~\bibnamefont {Morita}},\
  }\bibfield  {title} {\bibinfo {title} {{Proton-neutron and proton-proton
  correlations in medium-weight nuclei and the role of the tensor force}},\
  }\href {https://doi.org/10.1103/PhysRevLett.100.162503} {\bibfield  {journal}
  {\bibinfo  {journal} {Phys. Rev. Lett.}\ }\textbf {\bibinfo {volume} {100}},\
  \bibinfo {pages} {162503} (\bibinfo {year} {2008})}\BibitemShut {NoStop}%
%%CITATION = PRLTA,100,162503;%%
\bibitem [{\citenamefont {Ye}\ \emph {et~al.}(2018)\citenamefont {Ye} \emph
  {et~al.}}]{Ye18}%
  \BibitemOpen
  \bibfield  {author} {\bibinfo {author} {\bibfnamefont {Z.}~\bibnamefont {Ye}}
  \emph {et~al.} (\bibinfo {collaboration} {The Jefferson Lab Hall A
  Collaboration}),\ }\bibfield  {title} {\bibinfo {title} {Search for
  three-nucleon short-range correlations in light nuclei},\ }\href
  {https://doi.org/10.1103/PhysRevC.97.065204} {\bibfield  {journal} {\bibinfo
  {journal} {Phys. Rev. C}\ }\textbf {\bibinfo {volume} {97}},\ \bibinfo
  {pages} {065204} (\bibinfo {year} {2018})}\BibitemShut {NoStop}%
\bibitem [{\citenamefont {Sargsian}\ \emph {et~al.}(2019)\citenamefont
  {Sargsian}, \citenamefont {Day}, \citenamefont {Frankfurt},\ and\
  \citenamefont {Strikman}}]{Sargasian19}%
  \BibitemOpen
  \bibfield  {author} {\bibinfo {author} {\bibfnamefont {M.~M.}\ \bibnamefont
  {Sargsian}}, \bibinfo {author} {\bibfnamefont {D.~B.}\ \bibnamefont {Day}},
  \bibinfo {author} {\bibfnamefont {L.~L.}\ \bibnamefont {Frankfurt}},\ and\
  \bibinfo {author} {\bibfnamefont {M.~I.}\ \bibnamefont {Strikman}},\
  }\bibfield  {title} {\bibinfo {title} {Searching for three-nucleon
  short-range correlations},\ }\href
  {https://doi.org/10.1103/PhysRevC.100.044320} {\bibfield  {journal} {\bibinfo
   {journal} {Phys. Rev. C}\ }\textbf {\bibinfo {volume} {100}},\ \bibinfo
  {pages} {044320} (\bibinfo {year} {2019})}\BibitemShut {NoStop}%
\bibitem [{\citenamefont {Tan}(2008{\natexlab{a}})}]{Tan08a}%
  \BibitemOpen
  \bibfield  {author} {\bibinfo {author} {\bibfnamefont {S.}~\bibnamefont
  {Tan}},\ }\bibfield  {title} {\bibinfo {title} {Energetics of a strongly
  correlated fermi gas},\ }\href@noop {} {\bibfield  {journal} {\bibinfo
  {journal} {Annals of Physics}\ }\textbf {\bibinfo {volume} {323}},\ \bibinfo
  {pages} {2952} (\bibinfo {year} {2008}{\natexlab{a}})}\BibitemShut {NoStop}%
\bibitem [{\citenamefont {Tan}(2008{\natexlab{b}})}]{Tan08b}%
  \BibitemOpen
  \bibfield  {author} {\bibinfo {author} {\bibfnamefont {S.}~\bibnamefont
  {Tan}},\ }\bibfield  {title} {\bibinfo {title} {Large momentum part of a
  strongly correlated fermi gas},\ }\href
  {https://doi.org/http://dx.doi.org/10.1016/j.aop.2008.03.005} {\bibfield
  {journal} {\bibinfo  {journal} {Annals of Physics}\ }\textbf {\bibinfo
  {volume} {323}},\ \bibinfo {pages} {2971} (\bibinfo {year}
  {2008}{\natexlab{b}})}\BibitemShut {NoStop}%
\bibitem [{\citenamefont {Tan}(2008{\natexlab{c}})}]{Tan08c}%
  \BibitemOpen
  \bibfield  {author} {\bibinfo {author} {\bibfnamefont {S.}~\bibnamefont
  {Tan}},\ }\bibfield  {title} {\bibinfo {title} {Generalized virial theorem
  and pressure relation for a strongly correlated fermi gas},\ }\href
  {https://doi.org/http://dx.doi.org/10.1016/j.aop.2008.03.003} {\bibfield
  {journal} {\bibinfo  {journal} {Annals of Physics}\ }\textbf {\bibinfo
  {volume} {323}},\ \bibinfo {pages} {2987} (\bibinfo {year}
  {2008}{\natexlab{c}})}\BibitemShut {NoStop}%
\bibitem [{\citenamefont {Braaten}(2012)}]{Braaten12}%
  \BibitemOpen
  \bibfield  {author} {\bibinfo {author} {\bibfnamefont {E.}~\bibnamefont
  {Braaten}},\ }\bibfield  {title} {\bibinfo {title} {Universal relations for
  fermions with large scattering length},\ }in\ \href@noop {} {\emph {\bibinfo
  {booktitle} {The BCS-BEC Crossover and the Unitary Fermi Gas}}},\ \bibinfo
  {editor} {edited by\ \bibinfo {editor} {\bibfnamefont {W.}~\bibnamefont
  {Zwerger}}}\ (\bibinfo  {publisher} {Springer},\ \bibinfo {address}
  {Berlin},\ \bibinfo {year} {2012})\BibitemShut {NoStop}%
\bibitem [{\citenamefont {Weiss}\ \emph
  {et~al.}(2015{\natexlab{a}})\citenamefont {Weiss}, \citenamefont {Bazak},\
  and\ \citenamefont {Barnea}}]{Weiss14}%
  \BibitemOpen
  \bibfield  {author} {\bibinfo {author} {\bibfnamefont {R.}~\bibnamefont
  {Weiss}}, \bibinfo {author} {\bibfnamefont {B.}~\bibnamefont {Bazak}},\ and\
  \bibinfo {author} {\bibfnamefont {N.}~\bibnamefont {Barnea}},\ }\bibfield
  {title} {\bibinfo {title} {Nuclear neutron-proton contact and the
  photoabsorption cross section},\ }\href
  {https://doi.org/10.1103/PhysRevLett.114.012501} {\bibfield  {journal}
  {\bibinfo  {journal} {Phys. Rev. Lett.}\ }\textbf {\bibinfo {volume} {114}},\
  \bibinfo {pages} {012501} (\bibinfo {year} {2015}{\natexlab{a}})}\BibitemShut
  {NoStop}%
\bibitem [{\citenamefont {Weiss}\ \emph
  {et~al.}(2015{\natexlab{b}})\citenamefont {Weiss}, \citenamefont {Bazak},\
  and\ \citenamefont {Barnea}}]{Weiss:2015mba}%
  \BibitemOpen
  \bibfield  {author} {\bibinfo {author} {\bibfnamefont {R.}~\bibnamefont
  {Weiss}}, \bibinfo {author} {\bibfnamefont {B.}~\bibnamefont {Bazak}},\ and\
  \bibinfo {author} {\bibfnamefont {N.}~\bibnamefont {Barnea}},\ }\bibfield
  {title} {\bibinfo {title} {{Generalized nuclear contacts and momentum
  distributions}},\ }\href {https://doi.org/10.1103/PhysRevC.92.054311}
  {\bibfield  {journal} {\bibinfo  {journal} {Phys. Rev.}\ }\textbf {\bibinfo
  {volume} {C92}},\ \bibinfo {pages} {054311} (\bibinfo {year}
  {2015}{\natexlab{b}})},\ \Eprint {https://arxiv.org/abs/1503.07047}
  {arXiv:1503.07047 [nucl-th]} \BibitemShut {NoStop}%
%%CITATION = ARXIV:1503.07047;%%
\bibitem [{\citenamefont {Weiss}\ \emph
  {et~al.}(2016{\natexlab{a}})\citenamefont {Weiss}, \citenamefont {Pazy},\
  and\ \citenamefont {Barnea}}]{Weiss_2016}%
  \BibitemOpen
  \bibfield  {author} {\bibinfo {author} {\bibfnamefont {R.}~\bibnamefont
  {Weiss}}, \bibinfo {author} {\bibfnamefont {E.}~\bibnamefont {Pazy}},\ and\
  \bibinfo {author} {\bibfnamefont {N.}~\bibnamefont {Barnea}},\ }\bibfield
  {title} {\bibinfo {title} {Short range correlations: The important role of
  few-body dynamics in many-body systems},\ }\bibfield  {journal} {\bibinfo
  {journal} {Few-Body Systems}\ }\textbf {\bibinfo {volume} {58}},\ \href
  {https://doi.org/10.1007/s00601-016-1165-2} {10.1007/s00601-016-1165-2}
  (\bibinfo {year} {2016}{\natexlab{a}})\BibitemShut {NoStop}%
\bibitem [{\citenamefont {Weiss}\ and\ \citenamefont
  {Barnea}(2017)}]{Weiss17_CoupledChannels}%
  \BibitemOpen
  \bibfield  {author} {\bibinfo {author} {\bibfnamefont {R.}~\bibnamefont
  {Weiss}}\ and\ \bibinfo {author} {\bibfnamefont {N.}~\bibnamefont {Barnea}},\
  }\bibfield  {title} {\bibinfo {title} {Contact formalism for coupled
  channels},\ }\href {https://doi.org/10.1103/PhysRevC.96.041303} {\bibfield
  {journal} {\bibinfo  {journal} {Phys. Rev. C}\ }\textbf {\bibinfo {volume}
  {96}},\ \bibinfo {pages} {041303} (\bibinfo {year} {2017})}\BibitemShut
  {NoStop}%
\bibitem [{\citenamefont {Weiss}\ \emph {et~al.}(2018)\citenamefont {Weiss},
  \citenamefont {Cruz-Torres}, \citenamefont {Barnea}, \citenamefont
  {Piasetzky},\ and\ \citenamefont {Hen}}]{Weiss:2016obx}%
  \BibitemOpen
  \bibfield  {author} {\bibinfo {author} {\bibfnamefont {R.}~\bibnamefont
  {Weiss}}, \bibinfo {author} {\bibfnamefont {R.}~\bibnamefont {Cruz-Torres}},
  \bibinfo {author} {\bibfnamefont {N.}~\bibnamefont {Barnea}}, \bibinfo
  {author} {\bibfnamefont {E.}~\bibnamefont {Piasetzky}},\ and\ \bibinfo
  {author} {\bibfnamefont {O.}~\bibnamefont {Hen}},\ }\bibfield  {title}
  {\bibinfo {title} {{The nuclear contacts and short range correlations in
  nuclei}},\ }\href@noop {} {\bibfield  {journal} {\bibinfo  {journal} {Phys.
  Lett. B}\ }\textbf {\bibinfo {volume} {780}},\ \bibinfo {pages} {211}
  (\bibinfo {year} {2018})}\BibitemShut {NoStop}%
%%CITATION = ARXIV:1612.00923;%%
\bibitem [{\citenamefont {Cruz-Torres}\ \emph {et~al.}(2020)\citenamefont
  {Cruz-Torres}, \citenamefont {Lonardoni}, \citenamefont {Weiss},
  \citenamefont {Piarulli}, \citenamefont {Barnea}, \citenamefont
  {Higinbotham}, \citenamefont {Piasetzky}, \citenamefont {Schmidt},
  \citenamefont {Weinstein}, \citenamefont {Wiringa},\ and\ \citenamefont
  {Hen}}]{Cruz-Torres2020}%
  \BibitemOpen
  \bibfield  {author} {\bibinfo {author} {\bibfnamefont {R.}~\bibnamefont
  {Cruz-Torres}}, \bibinfo {author} {\bibfnamefont {D.}~\bibnamefont
  {Lonardoni}}, \bibinfo {author} {\bibfnamefont {R.}~\bibnamefont {Weiss}},
  \bibinfo {author} {\bibfnamefont {M.}~\bibnamefont {Piarulli}}, \bibinfo
  {author} {\bibfnamefont {N.}~\bibnamefont {Barnea}}, \bibinfo {author}
  {\bibfnamefont {D.~W.}\ \bibnamefont {Higinbotham}}, \bibinfo {author}
  {\bibfnamefont {E.}~\bibnamefont {Piasetzky}}, \bibinfo {author}
  {\bibfnamefont {A.}~\bibnamefont {Schmidt}}, \bibinfo {author} {\bibfnamefont
  {L.~B.}\ \bibnamefont {Weinstein}}, \bibinfo {author} {\bibfnamefont {R.~B.}\
  \bibnamefont {Wiringa}},\ and\ \bibinfo {author} {\bibfnamefont
  {O.}~\bibnamefont {Hen}},\ }\bibfield  {title} {\bibinfo {title} {{Many-body
  factorization and position-momentum equivalence of nuclear short-range
  correlations}},\ }\href {https://doi.org/10.1038/s41567-020-01053-7}
  {\bibfield  {journal} {\bibinfo  {journal} {Nature Physics}\ } (\bibinfo
  {year} {2020})},\ \Eprint {https://arxiv.org/abs/1907.03658}
  {arXiv:1907.03658 [nucl-th]} \BibitemShut {NoStop}%
%%CITATION = ARXIV:1907.03658;%%
\bibitem [{\citenamefont {Weiss}\ \emph
  {et~al.}(2016{\natexlab{b}})\citenamefont {Weiss}, \citenamefont {Bazak},\
  and\ \citenamefont {Barnea}}]{Weiss_EPJA16}%
  \BibitemOpen
  \bibfield  {author} {\bibinfo {author} {\bibfnamefont {R.}~\bibnamefont
  {Weiss}}, \bibinfo {author} {\bibfnamefont {B.}~\bibnamefont {Bazak}},\ and\
  \bibinfo {author} {\bibfnamefont {N.}~\bibnamefont {Barnea}},\ }\bibfield
  {title} {\bibinfo {title} {The generalized nuclear contact and its
  application to the photoabsorption cross-section},\ }\bibfield  {journal}
  {\bibinfo  {journal} {The European Physical Journal A}\ }\textbf {\bibinfo
  {volume} {52}},\ \href {https://doi.org/10.1140/epja/i2016-16092-3}
  {10.1140/epja/i2016-16092-3} (\bibinfo {year}
  {2016}{\natexlab{b}})\BibitemShut {NoStop}%
\bibitem [{\citenamefont {Weiss}\ \emph
  {et~al.}(2019{\natexlab{a}})\citenamefont {Weiss}, \citenamefont {Korover},
  \citenamefont {Piasetzky}, \citenamefont {Hen},\ and\ \citenamefont
  {Barnea}}]{Weiss:2018tbu}%
  \BibitemOpen
  \bibfield  {author} {\bibinfo {author} {\bibfnamefont {R.}~\bibnamefont
  {Weiss}}, \bibinfo {author} {\bibfnamefont {I.}~\bibnamefont {Korover}},
  \bibinfo {author} {\bibfnamefont {E.}~\bibnamefont {Piasetzky}}, \bibinfo
  {author} {\bibfnamefont {O.}~\bibnamefont {Hen}},\ and\ \bibinfo {author}
  {\bibfnamefont {N.}~\bibnamefont {Barnea}},\ }\bibfield  {title} {\bibinfo
  {title} {{Energy and momentum dependence of nuclear short-range correlations
  - Spectral function, exclusive scattering experiments and the contact
  formalism}},\ }\href {https://doi.org/10.1016/j.physletb.2019.02.019}
  {\bibfield  {journal} {\bibinfo  {journal} {Phys. Lett.}\ }\textbf {\bibinfo
  {volume} {B791}},\ \bibinfo {pages} {242} (\bibinfo {year}
  {2019}{\natexlab{a}})},\ \Eprint {https://arxiv.org/abs/1806.10217}
  {arXiv:1806.10217 [nucl-th]} \BibitemShut {NoStop}%
%%CITATION = ARXIV:1806.10217;%%
\bibitem [{\citenamefont {Weiss}\ \emph
  {et~al.}(2019{\natexlab{b}})\citenamefont {Weiss}, \citenamefont {Schmidt},
  \citenamefont {Miller},\ and\ \citenamefont {Barnea}}]{WEISS2019484}%
  \BibitemOpen
  \bibfield  {author} {\bibinfo {author} {\bibfnamefont {R.}~\bibnamefont
  {Weiss}}, \bibinfo {author} {\bibfnamefont {A.}~\bibnamefont {Schmidt}},
  \bibinfo {author} {\bibfnamefont {G.~A.}\ \bibnamefont {Miller}},\ and\
  \bibinfo {author} {\bibfnamefont {N.}~\bibnamefont {Barnea}},\ }\bibfield
  {title} {\bibinfo {title} {Short-range correlations and the charge density},\
  }\href {https://doi.org/https://doi.org/10.1016/j.physletb.2019.01.053}
  {\bibfield  {journal} {\bibinfo  {journal} {Physics Letters B}\ }\textbf
  {\bibinfo {volume} {790}},\ \bibinfo {pages} {484 } (\bibinfo {year}
  {2019}{\natexlab{b}})}\BibitemShut {NoStop}%
\bibitem [{\citenamefont {Pybus}\ \emph {et~al.}(2020)\citenamefont {Pybus},
  \citenamefont {Korover}, \citenamefont {Weiss}, \citenamefont {Schmidt},
  \citenamefont {Barnea}, \citenamefont {Higinbotham}, \citenamefont
  {Piasetzky}, \citenamefont {Strikman}, \citenamefont {Weinstein},\ and\
  \citenamefont {Hen}}]{Pybus:2020itv}%
  \BibitemOpen
  \bibfield  {author} {\bibinfo {author} {\bibfnamefont {J.}~\bibnamefont
  {Pybus}}, \bibinfo {author} {\bibfnamefont {I.}~\bibnamefont {Korover}},
  \bibinfo {author} {\bibfnamefont {R.}~\bibnamefont {Weiss}}, \bibinfo
  {author} {\bibfnamefont {A.}~\bibnamefont {Schmidt}}, \bibinfo {author}
  {\bibfnamefont {N.}~\bibnamefont {Barnea}}, \bibinfo {author} {\bibfnamefont
  {D.}~\bibnamefont {Higinbotham}}, \bibinfo {author} {\bibfnamefont
  {E.}~\bibnamefont {Piasetzky}}, \bibinfo {author} {\bibfnamefont
  {M.}~\bibnamefont {Strikman}}, \bibinfo {author} {\bibfnamefont
  {L.}~\bibnamefont {Weinstein}},\ and\ \bibinfo {author} {\bibfnamefont
  {O.}~\bibnamefont {Hen}},\ }\bibfield  {title} {\bibinfo {title}
  {{Generalized contact formalism analysis of the $^4$He$(e,e'pN)$ reaction}},\
  }\href {https://doi.org/10.1016/j.physletb.2020.135429} {\bibfield  {journal}
  {\bibinfo  {journal} {Phys. Lett. B}\ }\textbf {\bibinfo {volume} {805}},\
  \bibinfo {pages} {135429} (\bibinfo {year} {2020})},\ \Eprint
  {https://arxiv.org/abs/2003.02318} {arXiv:2003.02318 [nucl-th]} \BibitemShut
  {NoStop}%
\bibitem [{\citenamefont {Patsyuk}\ \emph {et~al.}(2021)\citenamefont
  {Patsyuk}, \citenamefont {Kahlbow}, \citenamefont {Laskaris}, \citenamefont
  {Duer}, \citenamefont {Lenivenko}, \citenamefont {Segarra}, \citenamefont
  {Atovullaev}, \citenamefont {Johansson}, \citenamefont {Aumann},
  \citenamefont {Corsi},\ and\ \citenamefont
  {et~al.}}]{patsyuk2021unperturbed}%
  \BibitemOpen
  \bibfield  {author} {\bibinfo {author} {\bibfnamefont {M.}~\bibnamefont
  {Patsyuk}}, \bibinfo {author} {\bibfnamefont {J.}~\bibnamefont {Kahlbow}},
  \bibinfo {author} {\bibfnamefont {G.}~\bibnamefont {Laskaris}}, \bibinfo
  {author} {\bibfnamefont {M.}~\bibnamefont {Duer}}, \bibinfo {author}
  {\bibfnamefont {V.}~\bibnamefont {Lenivenko}}, \bibinfo {author}
  {\bibfnamefont {E.~P.}\ \bibnamefont {Segarra}}, \bibinfo {author}
  {\bibfnamefont {T.}~\bibnamefont {Atovullaev}}, \bibinfo {author}
  {\bibfnamefont {G.}~\bibnamefont {Johansson}}, \bibinfo {author}
  {\bibfnamefont {T.}~\bibnamefont {Aumann}}, \bibinfo {author} {\bibfnamefont
  {A.}~\bibnamefont {Corsi}},\ and\ \bibinfo {author} {\bibnamefont {et~al.}},\
  }\bibfield  {title} {\bibinfo {title} {Unperturbed inverse kinematics nucleon
  knockout measurements with a carbon beam},\ }\href
  {https://doi.org/10.1038/s41567-021-01193-4} {\bibfield  {journal} {\bibinfo
  {journal} {Nature Physics}\ }\textbf {\bibinfo {volume} {17}},\ \bibinfo
  {pages} {693–699} (\bibinfo {year} {2021})}\BibitemShut {NoStop}%
\bibitem [{\citenamefont {Weiss}\ \emph
  {et~al.}(2021{\natexlab{a}})\citenamefont {Weiss}, \citenamefont {Denniston},
  \citenamefont {Pybus}, \citenamefont {Hen}, \citenamefont {Piasetzky},
  \citenamefont {Schmidt}, \citenamefont {Weinstein},\ and\ \citenamefont
  {Barnea}}]{weiss2020inclusive}%
  \BibitemOpen
  \bibfield  {author} {\bibinfo {author} {\bibfnamefont {R.}~\bibnamefont
  {Weiss}}, \bibinfo {author} {\bibfnamefont {A.~W.}\ \bibnamefont
  {Denniston}}, \bibinfo {author} {\bibfnamefont {J.~R.}\ \bibnamefont
  {Pybus}}, \bibinfo {author} {\bibfnamefont {O.}~\bibnamefont {Hen}}, \bibinfo
  {author} {\bibfnamefont {E.}~\bibnamefont {Piasetzky}}, \bibinfo {author}
  {\bibfnamefont {A.}~\bibnamefont {Schmidt}}, \bibinfo {author} {\bibfnamefont
  {L.~B.}\ \bibnamefont {Weinstein}},\ and\ \bibinfo {author} {\bibfnamefont
  {N.}~\bibnamefont {Barnea}},\ }\bibfield  {title} {\bibinfo {title}
  {Extracting the number of short-range correlated nucleon pairs from inclusive
  electron scattering data},\ }\href
  {https://doi.org/10.1103/PhysRevC.103.L031301} {\bibfield  {journal}
  {\bibinfo  {journal} {Phys. Rev. C}\ }\textbf {\bibinfo {volume} {103}},\
  \bibinfo {pages} {L031301} (\bibinfo {year}
  {2021}{\natexlab{a}})}\BibitemShut {NoStop}%
\bibitem [{\citenamefont {Weiss}\ \emph
  {et~al.}(2021{\natexlab{b}})\citenamefont {Weiss}, \citenamefont {Soriano},
  \citenamefont {Lovato}, \citenamefont {Menendez},\ and\ \citenamefont
  {Wiringa}}]{weiss2021neutrinoless}%
  \BibitemOpen
  \bibfield  {author} {\bibinfo {author} {\bibfnamefont {R.}~\bibnamefont
  {Weiss}}, \bibinfo {author} {\bibfnamefont {P.}~\bibnamefont {Soriano}},
  \bibinfo {author} {\bibfnamefont {A.}~\bibnamefont {Lovato}}, \bibinfo
  {author} {\bibfnamefont {J.}~\bibnamefont {Menendez}},\ and\ \bibinfo
  {author} {\bibfnamefont {R.~B.}\ \bibnamefont {Wiringa}},\ }\href@noop {}
  {\bibinfo {title} {Neutrinoless double-beta decay: combining quantum monte
  carlo and the nuclear shell model with the generalized contact formalism}}
  (\bibinfo {year} {2021}{\natexlab{b}}),\ \Eprint
  {https://arxiv.org/abs/2112.08146} {arXiv:2112.08146 [nucl-th]} \BibitemShut
  {NoStop}%
\bibitem [{\citenamefont {Anderson}\ \emph {et~al.}(2010)\citenamefont
  {Anderson}, \citenamefont {Bogner}, \citenamefont {Furnstahl},\ and\
  \citenamefont {Perry}}]{Anderson2010}%
  \BibitemOpen
  \bibfield  {author} {\bibinfo {author} {\bibfnamefont {E.~R.}\ \bibnamefont
  {Anderson}}, \bibinfo {author} {\bibfnamefont {S.~K.}\ \bibnamefont
  {Bogner}}, \bibinfo {author} {\bibfnamefont {R.~J.}\ \bibnamefont
  {Furnstahl}},\ and\ \bibinfo {author} {\bibfnamefont {R.~J.}\ \bibnamefont
  {Perry}},\ }\bibfield  {title} {\bibinfo {title} {Operator evolution via the
  similarity renormalization group: The deuteron},\ }\href
  {https://doi.org/10.1103/PhysRevC.82.054001} {\bibfield  {journal} {\bibinfo
  {journal} {Phys. Rev. C}\ }\textbf {\bibinfo {volume} {82}},\ \bibinfo
  {pages} {054001} (\bibinfo {year} {2010})}\BibitemShut {NoStop}%
\bibitem [{\citenamefont {Bogner}\ and\ \citenamefont
  {Roscher}(2012)}]{Bogner12}%
  \BibitemOpen
  \bibfield  {author} {\bibinfo {author} {\bibfnamefont {S.~K.}\ \bibnamefont
  {Bogner}}\ and\ \bibinfo {author} {\bibfnamefont {D.}~\bibnamefont
  {Roscher}},\ }\bibfield  {title} {\bibinfo {title} {High-momentum tails from
  low-momentum effective theories},\ }\href
  {https://doi.org/10.1103/PhysRevC.86.064304} {\bibfield  {journal} {\bibinfo
  {journal} {Phys. Rev. C}\ }\textbf {\bibinfo {volume} {86}},\ \bibinfo
  {pages} {064304} (\bibinfo {year} {2012})}\BibitemShut {NoStop}%
\bibitem [{\citenamefont {Tropiano}\ \emph {et~al.}(2021)\citenamefont
  {Tropiano}, \citenamefont {Bogner},\ and\ \citenamefont
  {Furnstahl}}]{Tropiano2021}%
  \BibitemOpen
  \bibfield  {author} {\bibinfo {author} {\bibfnamefont {A.~J.}\ \bibnamefont
  {Tropiano}}, \bibinfo {author} {\bibfnamefont {S.~K.}\ \bibnamefont
  {Bogner}},\ and\ \bibinfo {author} {\bibfnamefont {R.~J.}\ \bibnamefont
  {Furnstahl}},\ }\bibfield  {title} {\bibinfo {title} {Short-range correlation
  physics at low renormalization group resolution},\ }\href
  {https://doi.org/10.1103/PhysRevC.104.034311} {\bibfield  {journal} {\bibinfo
   {journal} {Phys. Rev. C}\ }\textbf {\bibinfo {volume} {104}},\ \bibinfo
  {pages} {034311} (\bibinfo {year} {2021})}\BibitemShut {NoStop}%
\bibitem [{\citenamefont {Bartlett}\ and\ \citenamefont
  {Musia\l{}}(2007)}]{Bartlett07}%
  \BibitemOpen
  \bibfield  {author} {\bibinfo {author} {\bibfnamefont {R.~J.}\ \bibnamefont
  {Bartlett}}\ and\ \bibinfo {author} {\bibfnamefont {M.}~\bibnamefont
  {Musia\l{}}},\ }\bibfield  {title} {\bibinfo {title} {Coupled-cluster theory
  in quantum chemistry},\ }\href {https://doi.org/10.1103/RevModPhys.79.291}
  {\bibfield  {journal} {\bibinfo  {journal} {Rev. Mod. Phys.}\ }\textbf
  {\bibinfo {volume} {79}},\ \bibinfo {pages} {291} (\bibinfo {year}
  {2007})}\BibitemShut {NoStop}%
\bibitem [{\citenamefont {Hagen}\ \emph
  {et~al.}(2014{\natexlab{a}})\citenamefont {Hagen}, \citenamefont
  {Papenbrock}, \citenamefont {Hjorth-Jensen},\ and\ \citenamefont
  {Dean}}]{Hagen_2014}%
  \BibitemOpen
  \bibfield  {author} {\bibinfo {author} {\bibfnamefont {G.}~\bibnamefont
  {Hagen}}, \bibinfo {author} {\bibfnamefont {T.}~\bibnamefont {Papenbrock}},
  \bibinfo {author} {\bibfnamefont {M.}~\bibnamefont {Hjorth-Jensen}},\ and\
  \bibinfo {author} {\bibfnamefont {D.~J.}\ \bibnamefont {Dean}},\ }\bibfield
  {title} {\bibinfo {title} {Coupled-cluster computations of atomic nuclei},\
  }\href {https://doi.org/10.1088/0034-4885/77/9/096302} {\bibfield  {journal}
  {\bibinfo  {journal} {Reports on Progress in Physics}\ }\textbf {\bibinfo
  {volume} {77}},\ \bibinfo {pages} {096302} (\bibinfo {year}
  {2014}{\natexlab{a}})}\BibitemShut {NoStop}%
\bibitem [{\citenamefont {I.~Shavitt}(2009)}]{shavit2009}%
  \BibitemOpen
  \bibfield  {author} {\bibinfo {author} {\bibfnamefont {R.~J.~B.}\
  \bibnamefont {I.~Shavitt}},\ }\href@noop {} {\emph {\bibinfo {title}
  {Many-Body Methods in Chemistry and Physics MBPT and Coupled-Cluster
  Theory}}}\ (\bibinfo  {publisher} {Cambridge Molecular Science},\ \bibinfo
  {year} {2009})\BibitemShut {NoStop}%
\bibitem [{\citenamefont {Baardsen}\ \emph {et~al.}(2013)\citenamefont
  {Baardsen}, \citenamefont {Ekstr\"om}, \citenamefont {Hagen},\ and\
  \citenamefont {Hjorth-Jensen}}]{CCNM2013}%
  \BibitemOpen
  \bibfield  {author} {\bibinfo {author} {\bibfnamefont {G.}~\bibnamefont
  {Baardsen}}, \bibinfo {author} {\bibfnamefont {A.}~\bibnamefont {Ekstr\"om}},
  \bibinfo {author} {\bibfnamefont {G.}~\bibnamefont {Hagen}},\ and\ \bibinfo
  {author} {\bibfnamefont {M.}~\bibnamefont {Hjorth-Jensen}},\ }\bibfield
  {title} {\bibinfo {title} {Coupled-cluster studies of infinite nuclear
  matter},\ }\href {https://doi.org/10.1103/PhysRevC.88.054312} {\bibfield
  {journal} {\bibinfo  {journal} {Phys. Rev. C}\ }\textbf {\bibinfo {volume}
  {88}},\ \bibinfo {pages} {054312} (\bibinfo {year} {2013})}\BibitemShut
  {NoStop}%
\bibitem [{\citenamefont {Hagen}\ \emph
  {et~al.}(2014{\natexlab{b}})\citenamefont {Hagen}, \citenamefont
  {Papenbrock}, \citenamefont {Ekstr\"om}, \citenamefont {Wendt}, \citenamefont
  {Baardsen}, \citenamefont {Gandolfi}, \citenamefont {Hjorth-Jensen},\ and\
  \citenamefont {Horowitz}}]{CCNM2014}%
  \BibitemOpen
  \bibfield  {author} {\bibinfo {author} {\bibfnamefont {G.}~\bibnamefont
  {Hagen}}, \bibinfo {author} {\bibfnamefont {T.}~\bibnamefont {Papenbrock}},
  \bibinfo {author} {\bibfnamefont {A.}~\bibnamefont {Ekstr\"om}}, \bibinfo
  {author} {\bibfnamefont {K.~A.}\ \bibnamefont {Wendt}}, \bibinfo {author}
  {\bibfnamefont {G.}~\bibnamefont {Baardsen}}, \bibinfo {author}
  {\bibfnamefont {S.}~\bibnamefont {Gandolfi}}, \bibinfo {author}
  {\bibfnamefont {M.}~\bibnamefont {Hjorth-Jensen}},\ and\ \bibinfo {author}
  {\bibfnamefont {C.~J.}\ \bibnamefont {Horowitz}},\ }\bibfield  {title}
  {\bibinfo {title} {Coupled-cluster calculations of nucleonic matter},\ }\href
  {https://doi.org/10.1103/PhysRevC.89.014319} {\bibfield  {journal} {\bibinfo
  {journal} {Phys. Rev. C}\ }\textbf {\bibinfo {volume} {89}},\ \bibinfo
  {pages} {014319} (\bibinfo {year} {2014}{\natexlab{b}})}\BibitemShut
  {NoStop}%
\bibitem [{\citenamefont {Bishop}\ and\ \citenamefont
  {L\"uhrmann}(1978)}]{Bishop78}%
  \BibitemOpen
  \bibfield  {author} {\bibinfo {author} {\bibfnamefont {R.~F.}\ \bibnamefont
  {Bishop}}\ and\ \bibinfo {author} {\bibfnamefont {K.~H.}\ \bibnamefont
  {L\"uhrmann}},\ }\bibfield  {title} {\bibinfo {title} {Electron correlations:
  I. ground-state results in the high-density regime},\ }\href
  {https://doi.org/10.1103/PhysRevB.17.3757} {\bibfield  {journal} {\bibinfo
  {journal} {Phys. Rev. B}\ }\textbf {\bibinfo {volume} {17}},\ \bibinfo
  {pages} {3757} (\bibinfo {year} {1978})}\BibitemShut {NoStop}%
\bibitem [{\citenamefont {Amado}\ and\ \citenamefont
  {Woloshyn}(1976)}]{amado76}%
  \BibitemOpen
  \bibfield  {author} {\bibinfo {author} {\bibfnamefont {R.}~\bibnamefont
  {Amado}}\ and\ \bibinfo {author} {\bibfnamefont {R.}~\bibnamefont
  {Woloshyn}},\ }\bibfield  {title} {\bibinfo {title} {Momentum distributions
  in the nucleus},\ }\href
  {https://doi.org/http://dx.doi.org/10.1016/0370-2693(76)90067-8} {\bibfield
  {journal} {\bibinfo  {journal} {Physics Letters B}\ }\textbf {\bibinfo
  {volume} {62}},\ \bibinfo {pages} {253 } (\bibinfo {year}
  {1976})}\BibitemShut {NoStop}%
\bibitem [{\citenamefont {Zabolitzky}\ and\ \citenamefont
  {Ey}(1978)}]{Zabolitzky78}%
  \BibitemOpen
  \bibfield  {author} {\bibinfo {author} {\bibfnamefont {J.}~\bibnamefont
  {Zabolitzky}}\ and\ \bibinfo {author} {\bibfnamefont {W.}~\bibnamefont
  {Ey}},\ }\bibfield  {title} {\bibinfo {title} {Momentum distributions of
  nucleons in nuclei},\ }\href
  {https://doi.org/https://doi.org/10.1016/0370-2693(78)90846-8} {\bibfield
  {journal} {\bibinfo  {journal} {Physics Letters B}\ }\textbf {\bibinfo
  {volume} {76}},\ \bibinfo {pages} {527 } (\bibinfo {year}
  {1978})}\BibitemShut {NoStop}%
\bibitem [{\citenamefont {Bloch}\ and\ \citenamefont {Horowitz}(1958)}]{BH58}%
  \BibitemOpen
  \bibfield  {author} {\bibinfo {author} {\bibfnamefont {C.}~\bibnamefont
  {Bloch}}\ and\ \bibinfo {author} {\bibfnamefont {J.}~\bibnamefont
  {Horowitz}},\ }\bibfield  {title} {\bibinfo {title} {Sur la détermination
  des premiers états d'un système de fermions dans le cas dégénéré},\
  }\href {https://doi.org/https://doi.org/10.1016/0029-5582(58)90136-6}
  {\bibfield  {journal} {\bibinfo  {journal} {Nuclear Physics}\ }\textbf
  {\bibinfo {volume} {8}},\ \bibinfo {pages} {91} (\bibinfo {year}
  {1958})}\BibitemShut {NoStop}%
\end{thebibliography}%

% \begin{thebibliography}{99}

% % Tan relations 
% \bibitem{Tan08}%
%   S. Tan,
%   Energetics of a strongly correlated Fermi gas,
%   Ann. Phys. (N.Y.) {\bf 323}, 2952 (2008);
%   Large momentum part of a strongly correlated Fermi gas,
%   Ann. Phys. (N.Y.) {\bf 323}, 2971 (2008);
%   Generalized virial theorem and pressure relation for a strongly correlated
%   Fermi gas,
%   Ann. Phys. (N.Y.) {\bf 323}, 2987 (2008). 

% \bibitem{HM99}
%  J. H. Heisenberg, and B. Mihaila, 
%  Ground state correlations and mean field in $^{16}$O,
%  Phys. Rev. C {\bf 59}, 1440 (1999).

% \bibitem{BM07}
%   R. J. Bartlett, and M. Musial,
%   Coupled-cluster theory in quantum chemistry,
%   Rev. Mod. Phys. {\bf 79}, 291 (2007).
  
% \bibitem{sh09}
%     I. Shavitt, R. J. Bartlett,
%     Many-Body Methods in Chemistry and Physics MBPT and Coupled-Cluster Theory,
%     Cambridge Molecular Science  2009
    
% \bibitem{gl83}
%     W. Gl{\" o}ckle,
%     The Quantum Mechanical Few-Body Problem,
%     Springer-Verlag 1983
% \end{thebibliography}
\end{document}